\documentclass[aps, prl, twocolumn,superscriptaddress,floatfix]{revtex4-1}
\usepackage[utf8]{inputenc}
\usepackage[T1]{fontenc}
\usepackage[greek, english]{babel}
\newcommand\koppa{\text{\selectlanguage{greek}{\qoppa}}}
\usepackage{multirow}
\usepackage{graphicx,amsfonts,amssymb,amsmath,mathrsfs,hyperref,bm,dsfont}
\usepackage{stackengine}
\usepackage{relsize}
\usepackage{comment}
\usepackage{physics}
\usepackage{braket}

\newif\ifhyper
\hypertrue
\ifhyper
\hypersetup{
   citecolor = {red},
   colorlinks = {true}, % false
   linkcolor = {blue},
   urlcolor = {blue} % magenta
}
\fi

\def\be{\begin{equation}}
\def\ee{\end{equation}}
\def\bea{\begin{eqnarray}}
\def\eea{\end{eqnarray}}

\begin{document}

\title{Continuously varying critical exponents in long-range quantum spin ladders}
\author{Patrick Adelhardt}
\affiliation{Friedrich-Alexander-Universit\"at Erlangen-N\"urnberg (FAU), Department Physik, Staudtstra{\ss}e 7, D-91058 Erlangen, Germany}
\author{Kai Phillip Schmidt}
\affiliation{Friedrich-Alexander-Universit\"at Erlangen-N\"urnberg (FAU), Department Physik, Staudtstra{\ss}e 7, D-91058 Erlangen, Germany}

\begin{abstract}
	We investigate the quantum-critical behavior between the rung-singlet phase with hidden string order and the N\'eel phase with broken $SU(2)$-symmetry on quantum spin ladders with algebraically decaying unfrustrated long-range Heisenberg interactions. Combining perturbative continuous unitary transformations (pCUT) with a white-graph expansion and Monte Carlo simulations yields high-order series expansions of energies and observables in the thermodynamic limit about the isolated rung-dimer limit. The breakdown of the rung-singlet phase allows to determine the critical line and the entire set of critical exponents as a function of the decay exponent of the long-range interaction. A non-trivial regime of continuously varying critical exponents as well as long-range mean-field behavior is demonstrated reminiscent of the long-range transverse-field Ising model.
\end{abstract}

\maketitle

%Introduction
%%%%%%%%%%%%%%%%%%%%%%%%%%%%%%%%%%%%%%%%%%%%%%%%%%%%%%%%%%%%%%%%%%%%%%%%%%%%%%%%%%%%%%%%%%%%
{\color{blue}\emph{Introduction.-}} 
While in electromagnetism the interaction between charged particles is long-range decaying as a power-law with distance, in condensed matter systems the interaction is typically screened, justifying to consider short-range interactions in most microscopic investigations. There are, however, notable examples where the long-range behavior persists like in conventional dipolar ferromagnetism \cite{Bitko1996, Chakraborty2004} and exotic spin-ice materials \cite{Bramwell2001, Castelnovo2008} giving rise to magnetic monopoles. In quantum optical platforms, long-range interactions are commonly present and there has been formidable experimental advancements over the past decades. Indeed, among others, ions in magneto-optical traps \cite{Islam2011, Britton2012, Islam2013, Jurcevic2014, Richerme2014, Mielenz2016, Bohnet2016, Jurcevic2017, Zhang2017, Zunkovic2018, Hempel2018, Joshi2022} and neutral atoms in optical lattices \cite{Weimer2010, Xia2015, Labuhn2016, Wang2016, Schauss2018, Leseleuc2019, Levine2019, Wu2019, Ebadi2021, Semeghini2021} have gained vast attention as these platforms can realize one- and two-dimensional lattices with adaptable geometries and a mesoscopic number of entities offering high-fidelity control and read-out. This makes them viable candidates for versatile quantum simulators and scalable quantum computers \cite{Saffman2010, Bruzewicz2019, Browaeys2020}. Both platforms realize effective Ising- and XY-type spin interactions which decay algebraically with distance. In neutral-atom platforms the decay exponent is fixed while it can be continuously tuned in trapped-ion systems. Recent progress ranges from the determination of molecular ground-state energies \cite{Hempel2018} and the realization of equilibrium \cite{Islam2011, Ebadi2021} and dynamical quantum phase transitions \cite{Zhang2017,Jurcevic2017, Zunkovic2018} to the direct observation of a symmetry-protected topological phase \cite{Leseleuc2019} or a topologically-ordered quantum spin liquid \cite{Semeghini2021} by measuring non-local string operators.\\\indent
The majority of numerical studies has focused on a variety of spin chains \cite{Sandvik2003, Koffel2012, Knap2013, Sun2017, Zhu2018, Sandvik2010, Vanderstraeten2018, Fey2016, Adelhardt2020, Langheld2022, Yang2020, Yusuf2004, Laflorencie2005, Zhu2006, Li2015, Tang2015, Gong2016, Maghrebi2017, Ren2020, Yang2020, Vodola2014, Vodola2015, Gong2016, Maity2019, Sadhukhan2021} as well as two-dimensional systems directly related to Rydberg atom platforms with quickly decaying ($\sim r^{-6}$) long-range interactions \cite{Samajdar2021, Verresen2021, Liu2022}. One prominent exception is the long-range transverse-field Ising model (LRTFIM), which was recently analyzed on the two-dimensional square and triangular lattice with tunable long-range interactions \cite{Humeniuk2016, Fey2019, Koziol2021}. Geometrically unfrustrated LRTFIMs in one and two dimensions are known to display three distinct regimes of quantum criticality between the high-field polarized phase and the low-field $\mathbb{Z}_2$-symmetry broken ground state: For short-range interactions the system exhibits nearest-neighbor criticality, for strong long-range interactions long-range mean-field behavior, and in-between continuously varying critical exponents \cite{Dutta2001, Sak1973, Defenu2017, Behan2017a, Behan2017b, Defenu2020}. \\\indent
Less is known about the quantum-critical behavior of systems with long-range interactions possessing a continuous symmetry like the antiferromagnetic spin-1/2 Heisenberg model. For the one-dimensional short-range Heisenberg chain, the spontaneous breaking of its continuous $SU(2)$-symmetry is forbidden by the Mermin-Wagner theorem \cite{Mermin1966, Hohenberg1967, Coleman1973, Bruno2001} and the system displays quasi long-range order with gapless fractional spinon excitations. Interestingly, this theorem can be circumvented when unfrustrated long-range interactions are sufficiently strong giving rise to a quantum phase transition to a N\'eel state with broken $SU(2)$-symmetry \cite{Yusuf2004, Laflorencie2005, Sandvik2010, Tang2015, Maghrebi2017, Yang2020, Yang2021}. Furthermore, a recent work \cite{Yang2022} has studied antiferromagnetic two-leg quantum spin ladders with unfrustrated long-range Heisenberg interactions. Here a quantum phase transition between the gapped short-range isotropic ladder and the N\'eel state with broken $SU(2)$-symmetry is present. Since the isotropic ladder displays a rung-singlet phase with non-local string-order parameter, a deconfined criticality \cite{Vishwanath2004, Senthil2004a, Senthil2004b, Senthil2005} between two distinct ordered quantum phases has been suggested. \\\indent
In this letter, we investigate two types of long-range quantum spin ladders with arbitrary ratios of nearest-neighbor leg and rung exchange coupling and for arbitrary decay exponent of the long-range Heisenberg interaction so that the system studied in Ref.~\cite{Yang2022} is contained as one specific parameter line. To this end we extend the pCUT approach developed in Ref.~\cite{Fey2019} to generic observables which allows us to locate the critical breakdown of the rung-singlet phase and determine the entire set of canonical critical exponents as a function of the decay exponent. A non-trivial regime of continuously varying critical exponents as well as long-range mean-field behavior is observed similar to the LRTFIM implying the absence of deconfined criticality. 

\begin{figure}[t]
	\includegraphics[width=0.9\columnwidth]{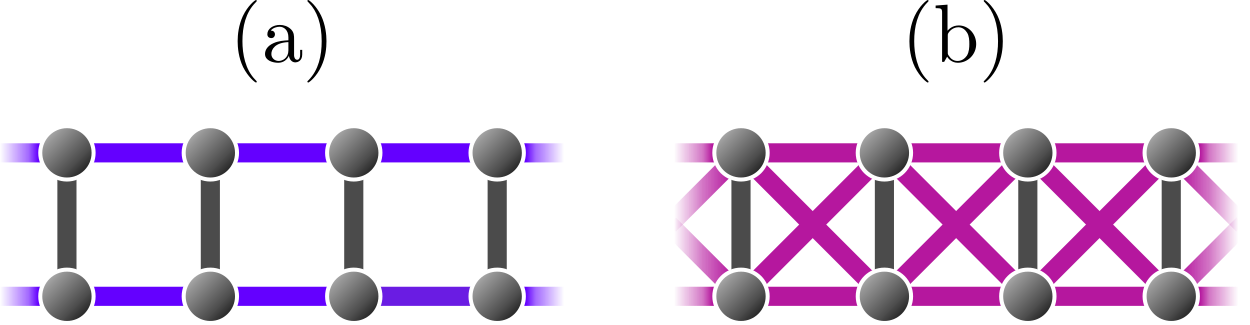}
	\caption{Illustration of the two underlying short-range quantum spin ladders with nearest-neighbor Heisenberg interaction $J_\perp$ on rung dimers as well as Heisenberg couplings between nearest-neighbor rung dimers (a) along the legs $\mathcal{H}_{\shortparallel}$ or (b) along the legs and the diagonals $\mathcal{H}_{\bowtie}$.}
	\label{fig:ladders}
\end{figure}

%Model
%%%%%%%%%%%%%%%%%%%%%%%%%%%%%%%%%%%%%%%%%%%%%%%%%%%%%%%%%%%%%%%%%%%%%%%%%%%%%%%%%%%%%%%%%%%%
{\color{blue}\emph{Model.-}}     
We consider the following spin-$1/2$ Hamiltonian
\begin{equation}
	\begin{split}
	\mathcal{H} = J_{\perp} &\sum_{i} \vec{S}_{i, 1} \vec{S}_{i, 2} - \sum_{i, \delta > 0} \sum_{n = 1}^{2} J_{\shortparallel}(\delta) \vec{S}_{i, n}\vec{S}_{i+\delta, n} \\ &- \sum_{i, \delta > 0} J_{\times}(\delta) \left(\vec{S}_{i, 1}\vec{S}_{i+\delta, 2} + \vec{S}_{i, 2}\vec{S}_{i+\delta, 1} \right),
	\end{split}
	\label{eq:HamSpin}
\end{equation}
where the indices $i$ and $i+\delta$ denote the rung position, the second index $n \in \{1, 2\}$ the leg of the ladder and the exchange parameters $J_{\perp}>0$, $J_{\shortparallel}(\delta)$, and $J_{\times}(\delta)$ couple spin operators on the rungs, legs, and diagonals, respectively. In the following, we restrict to the limiting cases $\mathcal{H}_{\shortparallel} \equiv \mathcal{H}\vert_{J_{\times}=0}$ and $\mathcal{H}_{\bowtie} \equiv \mathcal{H}\vert_{J_{\shortparallel} = J_{\times}}$, set $J_\perp=1$, and introduce $\lambda = J$ with $J \equiv J_{\shortparallel}$. Further, we define the distant-dependent coupling parameters to be of the form
\begin{equation}
\begin{gathered}
\begin{aligned}
&\lambda_{\shortparallel}(\delta) = \lambda\frac{(-1)^{\delta}}{\vert \delta |^{1+\sigma}}, \quad &\lambda_{\times}(\delta) = \lambda\frac{(-1)^{1+\delta}}{\vert 1+\delta|^{1+\sigma}},
\end{aligned}
\end{gathered}
\label{eq:Coupling}
\end{equation}
with $\lambda\geq 0$ realizing an unfrustrated algebraically decaying long-range interaction which induces antiferromagnetic Néel ordering for sufficiently small $\sigma$. Indeed, the decay exponent $\sigma$ can be tuned between the limiting cases of all-to-all interactions at $\sigma = -1$ and nearest-rung couplings at $\sigma = \infty$ (see Fig.~\ref{fig:ladders} for an illustration of the two different ladder models $\mathcal{H}_{\shortparallel}$ and $\mathcal{H}_{\bowtie}$ in the neighboring rung limit). Here, we focus on $\sigma\geq 0$ so that the energy of the system is extensive in the thermodynamic limit.
%Figure 2: Heisenberg ladders
%%%%%%%%%%%%%%%%%%%%%%%%%%%%%%%%%%%%%%%%%%%%%%%%%%%%%%%%%%%%%%%%%%%%%%%%%%%%%%%%%%%%%%%%%%%%
%%%%%%%%%%%%%%%%%%%%%%%%%%%%%%%%%%%%%%%%%%%%%%%%%%%%%%%%%%%%%%%%%%%%%%%%%%%%%%%%%%%%%%%%%%%%

In the limit of isolated rung dimers $\lambda=0$, the ground state is given exactly by the product state of rung singlets \mbox{$\ket{s}=~\left(\ket{\uparrow \downarrow} - \ket{\downarrow \uparrow}\right)/\sqrt{2}$} with localized rung triplets \mbox{$\ket{t_{x}}=-(\ket{\uparrow\uparrow} - \ket{\downarrow\downarrow})/\sqrt{2}$}, \mbox{$\ket{t_{y}}={\rm i}(\ket{\uparrow\uparrow}+\ket{\downarrow\downarrow})/\sqrt{2}$}, and $\ket{t_{z}}=\left(\ket{\uparrow\downarrow} + \ket{\downarrow\uparrow}\right)/\sqrt{2}$ as elementary excitations. For small $\lambda$ the ground state is adiabatically connected to this product state and the system is in the rung-singlet phase. The associated elementary excitations of the rung-singlet phase are gapped triplons corresponding to dressed rung-triplet excitations. For $\mathcal{H}_{\shortparallel}$ and $\sigma= \infty$ this holds for any finite $\lambda$ and only at $\lambda = \infty$ the system decouples into two spin-1/2 Heisenberg chains with gapless spinon excitations and a quasi long-range ordered ground state. In contrast, the Hamiltonian $\mathcal{H}_{\bowtie}$ becomes a diagonal spin-ladder at $\lambda=\infty$ for $\sigma= \infty$ and is expected to realize the gapped Haldane phase with exponentially decaying correlations for $\lambda>2$ falling into the category of symmetry-protected topological phases \cite{Kim2000, Pollmann2012}. The ground states of both Heisenberg ladders at any finite $\lambda$ only break a hidden $\mathbb{Z}_2\times \mathbb{Z}_2$ symmetry and can be characterized by non-local string order parameters \cite{Takada1992, Watanabe1995, Nishiyama1995, White1996, Kim2000}. \\\indent
Previous studies of the spin-$1/2$ Heisenberg chain \cite{Yusuf2004, Laflorencie2005, Tang2015, Yang2020} and the two-leg ladder $\mathcal{H}_{\bowtie}$ for $\lambda=1$ \cite{Yang2022} with unfrustrated long-range interactions  deduced a quantum phase transition towards Néel order with broken $SU(2)$-symmetry and thus circumventing the Mermin-Wagner theorem \cite{Mermin1966, Hohenberg1967, Coleman1973}. Further, Goldstone's theorem states that the spontaneous breaking of a continuous symmetry gives rise to massless Nambu-Goldstone modes \cite{Nambu1960, Goldstone1961, Goldstone1962}, however, the same restriction applies and the theorem loses its validity in the presence of long-range interactions. Indeed, in the extreme case of an all-to-all coupling the elementary excitation above the superextensive ground-state energy becomes proportional to the system size, evidently breaking Goldstone's theorem \cite{Yusuf2004, Yang2022}.

%Linked cluster expansions
%%%%%%%%%%%%%%%%%%%%%%%%%%%%%%%%%%%%%%%%%%%%%%%%%%%%%%%%%%%%%%%%%%%%%%%%%%%%%%%%%%%%%%%%%%%%
{\color{blue}\emph{pCUT method.-}} 
Interpreting the Heisenberg ladders as systems of coupled dimers, we introduce hard-core bosonic triplet (creation) annihilation operators $t_{i, \rho}^{(\dagger)}$ (creating) annihilating local triplets with flavor $\rho \in \{x, y, z\}$ on rung $i$ \cite{Sachdev1990, Hoermann2018}, the Hamiltonian \eqref{eq:HamSpin} can be written as
\begin{equation}
\mathcal{H} = \mathcal{H}_0 + \mathcal{V} = E_0 + \mathcal{Q} + \sum_{\delta>0}\lambda(\delta)\sum_{m=-2}^{2}T_m,
\label{eq:PCUTPrerequisite}
\end{equation}
where the unperturbed Hamiltonian of decoupled rungs is $\mathcal{H}_0 = E_0 + \mathcal{Q}$ and $E_0 = -3/4 N_{\rm rung}$ is the unperturbed ground-state energy with $N_{\rm rung}$ the number of rungs,  $\mathcal{Q} = \sum_{i, \rho} t_{i, \rho}^{\dagger}t_{i, \rho}^{\phantom\dagger}$ counts the number of triplet quasi-particles (QPs), and the perturbation $\mathcal{V}$ decomposes into a sum of operators $T_{m}$ containing processes of triplet operators that change the system's energy by $m\in\{0, \pm 2\}$ quanta. In the following we employ high-order series expansions along the same lines as in previous studies on the LRTFIM \cite{Fey2019, Koziol2019, Adelhardt2020, Langheld2022}. The pCUT method \cite{Knetter2000,Knetter2003} transforms the original Hamiltonian $\mathcal{H}$, perturbatively order by order in $\lambda$, into an effective Hamiltonian $\mathcal{H}_{\text{eff}}$ which conserves the number of triplon excitations $[\mathcal{Q}, \mathcal{H}_{\text{eff}}]= 0$. Similarly, observables $\mathcal{O}$ can be mapped to effective observables $\mathcal{O}_{\text{eff}}$ resulting in an expression analogous to $\mathcal{H}_{\text{eff}}$. However, the quasiparticle-conserving property is lost \cite{Knetter2003}. In a next step the effective Hamiltonian and observables have to be normal-ordered which is most efficiently done via a full-graph decomposition exploiting the linked-cluster theorem \cite{Coester2015}. For long-range interactions this is only feasible by applying a white-graph expansion \cite{Coester2015, Fey2016}, i.e.,~the most general linked contribution of a graph is extracted. To obtain the physical properties in the thermodynamic limit these general white-graph contributions have to be embedded on an infinite chain with rung dimers as effective supersites. Due to the presence of long-range interactions, every realization of a graph on the lattice except for overlapping configurations is possible leading to infinite nested sums that are evaluated using Markov-chain Monte Carlo integration for a fixed decay exponent $\sigma$ \cite{Fey2019} (see also Ref.~\cite{Suppl} for details). \\\indent
After applying the above procedure and Fourier transforming into quasi-momentum space, the effective one-triplon (1QP) Hamiltonian reads
\begin{equation}
	\tilde{\mathcal{H}}_{\text{eff}}^{\text{1QP}} =  \bar{E}_0 + \sum_{k, \rho} \omega(k)t^{\dagger}_{k, \rho}t_{k, \rho}^{\phantom{\dagger}},
\end{equation}
with the ground-state energy $\bar{E}_0$ and the 1QP dispersion $\omega(k)$. We can calculate the control-parameter susceptibility and the elementary one-triplon excitation gap directly
\begin{equation}
\begin{gathered}
\begin{aligned}
&\chi = \dv[2]{\bar{E}_0}{\lambda}, \qquad &\Delta = \min_{k}\omega(k) = \omega(k_{\rm c})
\end{aligned}
\end{gathered}
\end{equation}
with the critical momentum $k_{\rm c}=\pi$ for antiferromagnetic interactions. We further determine the one-triplon spectral weight. With $\mathcal{O}_{\rho}(k) = \frac{1}{2}(t_{k,\rho}^{\dagger} + t_{k, \rho}^{\phantom \dagger})$, one finds
\begin{equation}
\mathcal{S}^{\rm 1QP}(k) = \left| \bra{t_{k, \rho}} \mathcal{O}^{\text{1QP}}_{\text{eff}, \rho}(k) \ket{\rm ref} \right|^2 = |s(k)|^2,
\label{eq:ssw}
\end{equation}
where $\ket{\text{ref}} = \bigotimes_{i}\ket{s_i}$ is the unperturbed rung-singlet ground state, $\ket{t_{k, \rho}}$ is the one-triplon state with momentum $k$ and flavor $\rho \in \{x, y, z\}$, and $\mathcal{O}^{\text{1QP}}_{\text{eff}, \rho}(k) = s(k) (t_{k,\rho}^{\dagger} + t_{k, \rho}^{\phantom \dagger})$ the effective observable in second quantization restricted to the one-triplon channel. Here, we calculated the perturbative series of the ground-state energy $\bar{E}_0$ up to order 12 (8), the elementary gap $\Delta$ up to order 10 (7), and the one-triplon spectral weight $\mathcal{S}^{\rm 1QP}(k_{c})$ up to order 9 (7) in the perturbation parameter $\lambda$ for $\mathcal{H}_{\shortparallel}$ ($\mathcal{H}_{\bowtie}$). \\\indent
The above quantities show the dominant power-law behavior 
\begin{align}
\chi &\sim |\lambda - \lambda_{\rm c}|^{-\alpha}, \label{eq:powerlaw_chi}\\ 
\Delta &\sim |\lambda - \lambda_{\rm c}|^{z\nu}, \label{eq:powerlaw_delta}\\ 
\mathcal{S}^{\rm 1QP}(k_{\rm c}) &\sim |\lambda - \lambda_{\rm c}|^{-(2-z-\eta)\nu} \label{eq:powerlaw_sw}
\end{align}
close to the critical point $\lambda_c$ when the rung-singlet phase breaks down. The critical point and associated critical exponents can be directly determined from physical poles and associated residuals using (biased) DlogPadé extrapolants. More detailed information on the employed procedure from the graph embedding to extrapolations can be found in Ref.~\cite{Suppl}.

\begin{figure}[t]
	\includegraphics[width=\columnwidth]{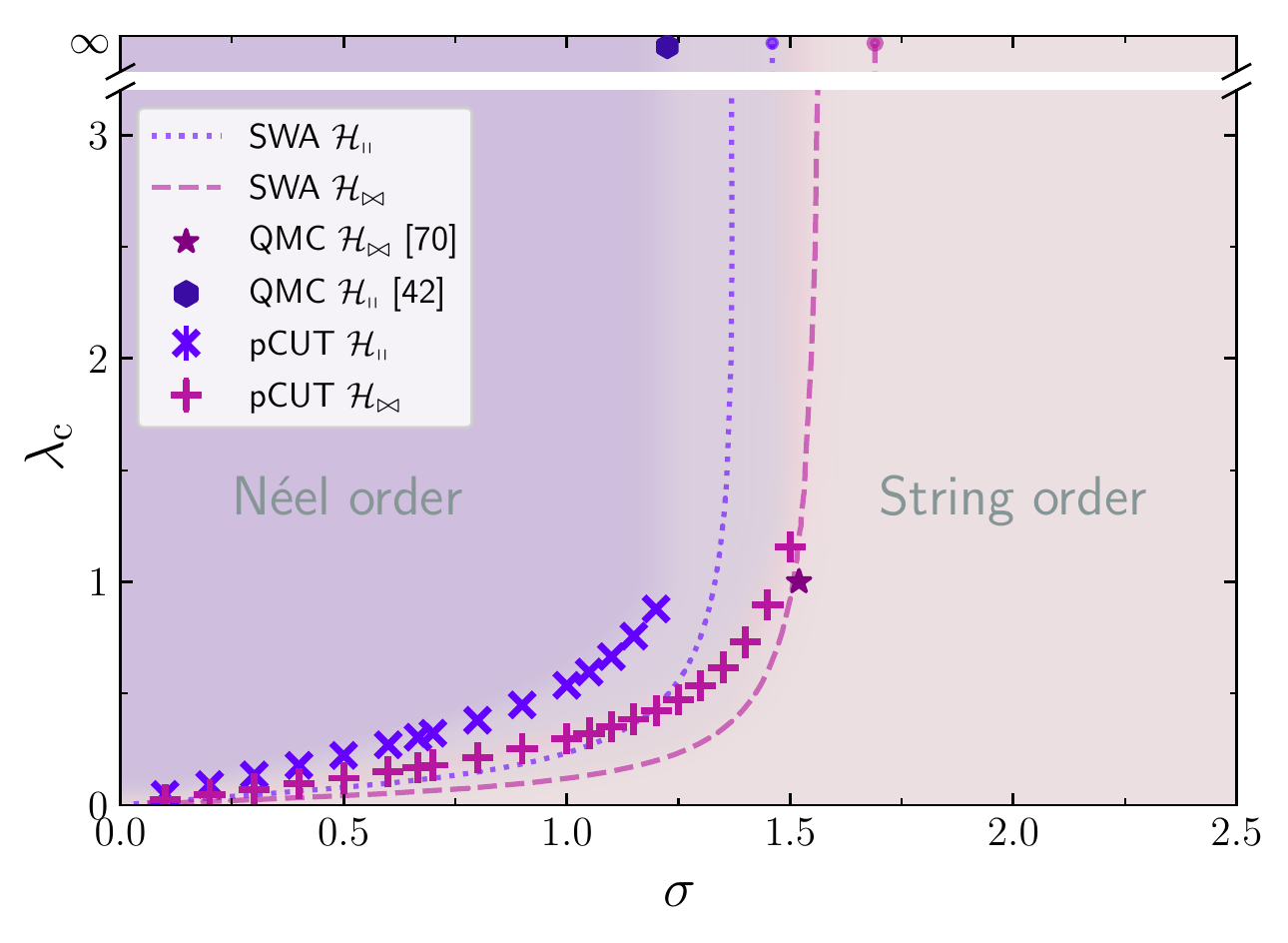}
	\caption{Phase diagram depicting the critical point $\lambda_c$ as a function of the decay exponent $\sigma$. Crosses are determined by DlogPadé extrapolations of the one-triplon gap series from the pCUT method while dashed lines are extracted from the self-consistency condition for the magnetization within linear spin-wave approximation (SWA). Diagonal crosses in navy blue correspond to $\mathcal{H}_{\shortparallel}$ and upright crosses in purple to $\mathcal{H}_{\bowtie}$. For the latter we observe that the Néel ordered phase sets in at smaller $\lambda$ or larger $\sigma$ exponents. The hexagon point at $\lambda=\infty$ for $\mathcal{H}_{\shortparallel}$ from Ref.~\cite{Laflorencie2005} as well the star-shaped point along the $\lambda=1$ line for $\mathcal{H}_{\bowtie}$ from Ref.~\cite{Yang2022} are consistent with our results.}
	\label{Fig:phase-diag}
\end{figure}

%Phase diagram
%%%%%%%%%%%%%%%%%%%%%%%%%%%%%%%%%%%%%%%%%%%%%%%%%%%%%%%%%%%%%%%%%%%%%%%%%%%%%%%%%%%%%%%%%%%%
{\color{blue}\emph{Quantum phase diagram.-}} 
We determine the phase transition point $\lambda_{\rm c}$ as a function of the decay exponent $\sigma$ by the quantum-critical breakdown of the rung-singlet phase and the accompanied closing of the elementary gap. The corresponding quantum phase diagram is shown in Fig.~\ref{Fig:phase-diag} for $\mathcal{H}_{\shortparallel}$ and $\mathcal{H}_{\bowtie}$. In the limit of $\lambda=0$ the system's ground state is given by the product of rung singlets and for $\sigma>2$ a quantum phase transition  can be ruled out from one-loop RG \cite{Dutta2001} since the $N$-component quantum rotor model can be mapped to the low-energy physics of the Heisenberg ladder \cite{Sachdev2011} in accordance with the Mermin-Wagner theorem.
%Figure 2: Phase diagram
%%%%%%%%%%%%%%%%%%%%%%%%%%%%%%%%%%%%%%%%%%%%%%%%%%%%%%%%%%%%%%%%%%%%%%%%%%%%%%%%%%%%%%%%%%%%
%%%%%%%%%%%%%%%%%%%%%%%%%%%%%%%%%%%%%%%%%%%%%%%%%%%%%%%%%%%%%%%%%%%%%%%%%%%%%%%%%%%%%%%%%%%%
At small $\sigma \lesssim 0.7$ ($\sigma \lesssim 1.0$) for $\mathcal{H}_{\shortparallel}$ ($\mathcal{H}_{\bowtie}$) the critical point $\lambda_c$ shifts linearly towards larger $\lambda$ with increasing $\sigma$. The gap closes earlier for $\mathcal{H}_{\bowtie}$ in agreement with expectation since the additional diagonal interactions further stabilize the antiferromagnetic N\'eel order. For larger $\sigma$ the critical points start to deviate from the linear behavior and bend upwards towards larger critical points until eventually DlogPadé extrapolations break down when the critical point shifts away significantly from the radius of convergence of the series. \\\indent
We complement the pCUT aproach with linear spin-wave calculations similar to \cite{Yusuf2004, Laflorencie2005}. Exploiting the fact that spin-wave theory is expected to work in the N\'eel ordered phase we can to determine the quantum-critical point (see Ref.~\cite{Suppl}). The spin-wave results are in qualitative rather than quantitative agreement with pCUT which is no surprise considering the crude approximation. In contrast to the pCUT results, the linear spin-wave calculations can be performed up to the limit $\lambda=\infty$. We find that $\lambda_{\rm c}$ diverges inline with expectations as the absence of criticality at large enough $\sigma$ suggests the existence of an upper critical decay exponent $\sigma_{\rm c}$. In fact, for $\mathcal{H}_{\shortparallel}$ at $\lambda=\infty$ we recover the spin-wave dispersion in Ref.~\cite{Laflorencie2005} yielding $\sigma^{\rm SW}_{\rm c}\approx 1.46$ and for $\mathcal{H}_{\bowtie}$ we find $\sigma^{\rm SW}_{\rm c}\approx 1.69$. Moreover, our data is consistent with $\sigma_c=1.225(25)$ from Ref.~\cite{Laflorencie2005} for $\mathcal{H}_{\shortparallel}$ and with $\sigma \approx 1.52$ at $\lambda_{\rm c}=1$ for $\mathcal{H}_{\bowtie}$ in Ref.~\cite{Yang2022} as depicted in Fig.~\ref{Fig:phase-diag}.

\begin{figure}[t]
	\includegraphics[width=\columnwidth]{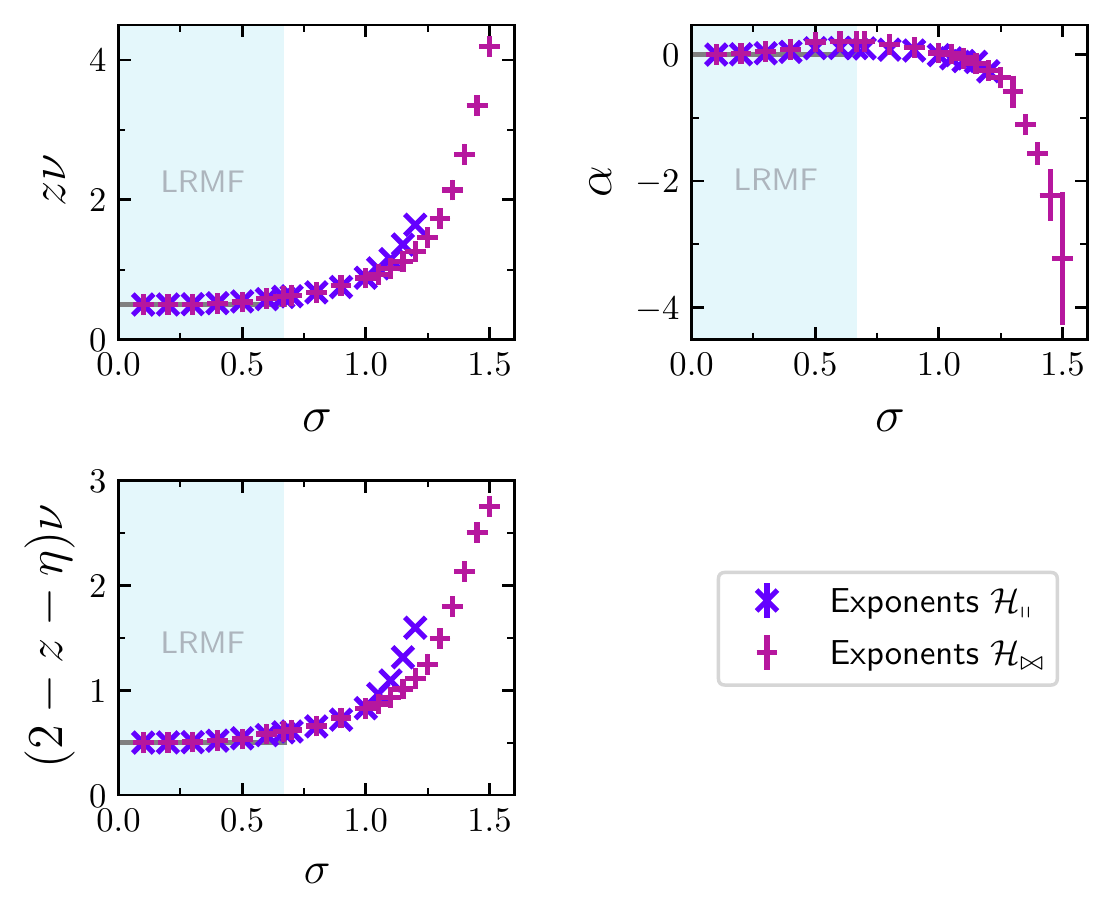}
	\caption{Critical exponents from Eqs.~\eqref{eq:powerlaw_chi}-\eqref{eq:powerlaw_sw} determined by the pCUT approach as a function of the decay exponent $\sigma$ for both ladder models $\mathcal{H}_{\shortparallel}$ and $\mathcal{H}_{\bowtie}$. For $\sigma\le 2/3$ the exponents clearly coincide with the expected long-range mean-field values (shaded region). For $\sigma>2/3$ they become continuously larger and apparently start to diverge. While the critical exponent for both models match well for $\sigma \lesssim 2.1$, they start to deviate from each other for larger values of $\sigma$ but this can probably be attributed to the difference in $\sigma_{\rm c}$.}
	\label{fig:exps_quantities}
\end{figure}

\begin{figure*}[t!]
	\centerline{\includegraphics[width=17cm]{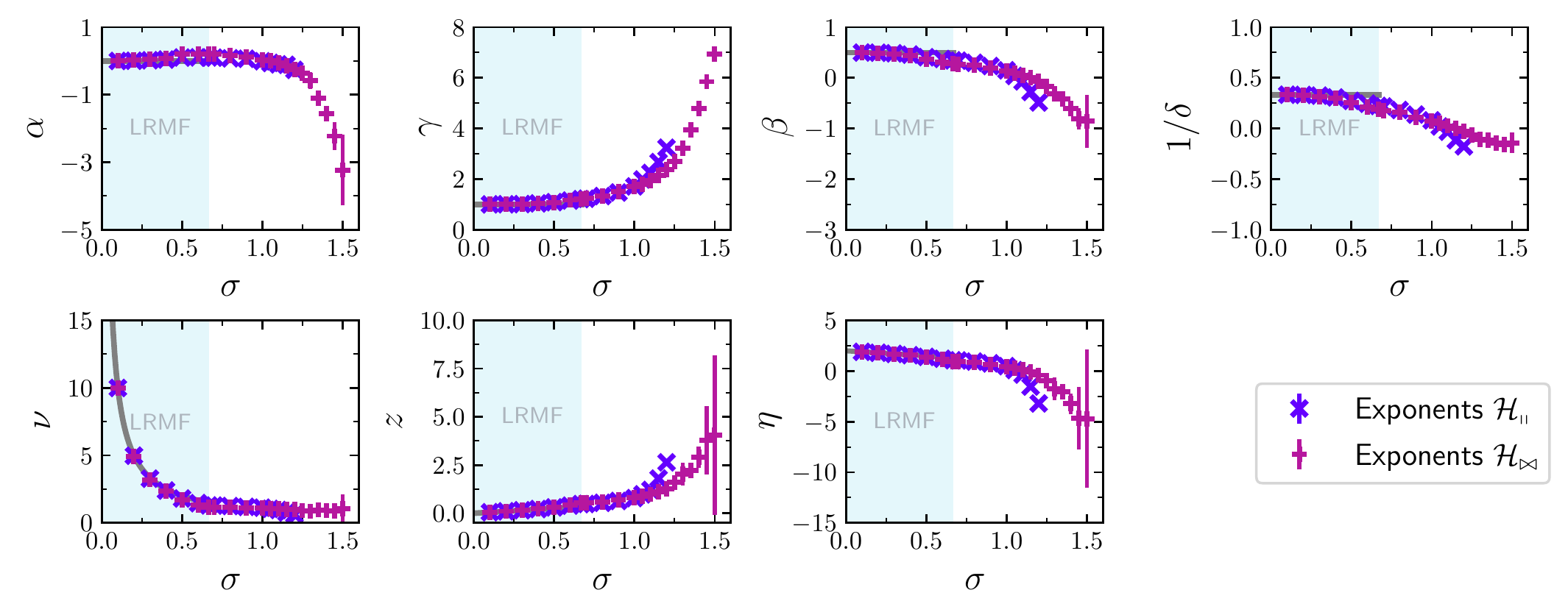}}
	\caption{Canonical critical exponents obtained from (hyper-) scaling relations as a function of the decay exponent $\sigma$. The critical exponents are in good agreement with expectations in the long-range mean-field regime (shaded region) and show continuously varying exponents for $\sigma> 2/3$. While some critical exponents appear to diverge others seem to go to a constant value for increasing $\sigma$. For some exponents the error bars become large for $\sigma\approx\sigma_c$.}
	\label{fig:exps_canonical}
\end{figure*}

%Critical Exponents
%%%%%%%%%%%%%%%%%%%%%%%%%%%%%%%%%%%%%%%%%%%%%%%%%%%%%%%%%%%%%%%%%%%%%%%%%%%%%%%%%%%%%%%%%%%%
{\color{blue}\emph{Critical Exponents.-}}
We extract the critical exponents according to Eqs.~\eqref{eq:powerlaw_chi}-\eqref{eq:powerlaw_sw} from DlogPadé extrapolants of the perturbative series. The exponents are depicted in Fig.~\ref{fig:exps_quantities} as a function of the decay exponent $\sigma$.
%Figure 3: Directly extracted critical exponents
%%%%%%%%%%%%%%%%%%%%%%%%%%%%%%%%%%%%%%%%%%%%%%%%%%%%%%%%%%%%%%%%%%%%%%%%%%%%%%%%%%%%%%%%%%%%
%%%%%%%%%%%%%%%%%%%%%%%%%%%%%%%%%%%%%%%%%%%%%%%%%%%%%%%%%%%%%%%%%%%%%%%%%%%%%%%%%%%%%%%%%%%%
The long-range mean-field regime (LRMF) is expected to extend to $\sigma \le 2/3$ \cite{Dutta2001}. The exponents extracted from DlogPadé extrapolants agree well with expected long-range mean-field exponents, although the presence of multiplicative logarithmic corrections to the dominant power-law behavior at the upper critical dimension $d_{\text{uc}}=2/(3\sigma)$ spoil the critical exponents.
%%%%%%%%%%%%%%%%%%%%%%%%%%%%%%%%%%%%%%%%%%%%%%%%%%%%%%%%%%%%%%%%%%%%%%%%%%%%%%%%%%%%%%%%%%%%
%Estimates for multiplicative logarithmic critical exponents can be found in Ref.~\cite{Suppl}.
%%%%%%%%%%%%%%%%%%%%%%%%%%%%%%%%%%%%%%%%%%%%%%%%%%%%%%%%%%%%%%%%%%%%%%%%%%%%%%%%%%%%%%%%%%%% 
Excluding the $\alpha$-exponent the critical exponents deviate less than 1.1\,\% (1.3\,\%) deep in the long-range regime $\sigma \le 0.3$ for $\mathcal{H}_{\shortparallel}$ ($\mathcal{H}_{\bowtie}$). We also observe continuously varying exponents for $\sigma > 2/3$ which seem to diverge for $\sigma\rightarrow\sigma_{\rm c}$. In terms of the gap closing this can be understood from the nearest-neighbor limit where the gap does not close but with the presence of long-range interactions the finite gap is lowered until eventually the gap closes. Further strengthening the long-range interactions shifts the critical point from infinity to smaller values and thus continuously tuning the exponent $z\nu$ from infinity to smaller values as the gap closes increasingly steep. In the region $\sigma \gtrsim 2.1$ for $\mathcal{H}_{\shortparallel}$ ($\sigma \gtrsim 2.2$ for $\mathcal{H}_{\bowtie}$) close to $\sigma_c$ it becomes difficult to extrapolate the critical point that starts to shift quickly towards $\lambda=\infty$ and therefore negatively affects the accuracy of the exponent estimates. \\\indent 
Using the three critical exponents from Fig.~\ref{fig:exps_quantities}, one can apply (hyper-) scaling relations to derive all canonical critical exponents (see Ref.~\cite{Suppl} for details) which are displayed in Fig.~\ref{fig:exps_canonical} for $\mathcal{H}_{\shortparallel}$ and $\mathcal{H}_{\bowtie}$. 
%Figure 4: Canonical critical exponents
%%%%%%%%%%%%%%%%%%%%%%%%%%%%%%%%%%%%%%%%%%%%%%%%%%%%%%%%%%%%%%%%%%%%%%%%%%%%%%%%%%%%%%%%%%%%
%%%%%%%%%%%%%%%%%%%%%%%%%%%%%%%%%%%%%%%%%%%%%%%%%%%%%%%%%%%%%%%%%%%%%%%%%%%%%%%%%%%%%%%%%%%%
In the long-range mean-field regime the exponents agree well with the expectations except for the exponents $\beta$ and $1/\delta$ around the upper critical dimension which we attribute to error propagation due to the presence of multiplicative logarithmic corrections. While the critical exponent $\gamma$ diverges for larger values of $\sigma$, the critical exponent $\nu$ approaches a constant value $\nu\approx 1$. The exponent $1/\delta$ goes to $-0.125$ in this limit, however we attribute this to a systematic error and that the correct limit might be $0$ since a jump of $\delta$ from $\infty$ to $-\infty$ is considered as unlikely. For the exponents $\beta$, $z$ and $\eta$ the uncertainty in the regime $\sigma \gtrsim 2.2$  becomes large due to error propagation and it is hard to make reliable statements in this region. \\\indent
Comparing the above results with Ref.~\cite{Yang2022} for $\mathcal{H}_{\bowtie}$ at $\lambda=1$ we find that the exponent $\nu= 1.8$ at about $\sigma\approx1.5$ is inconsistent with our result with constant $\nu=0.97(7)$ for all $\sigma>1.0$. Furthermore, the monotonously increasing exponent $z>1$ for $\sigma > 1.1$ is not in line with a proposed deconfined critical point with $z=1$ at $\sigma\approx1.5$. Our finding of continuously varying exponents reminiscent of the criticality of the unfrustrated LRTFIM \cite{Dutta2001, Sak1973, Defenu2017, Behan2017a, Behan2017b, Defenu2020, Fey2016, Fey2019} rises the question why this specific point should be extraordinary, particularly considering that despite the presence of a non-local string order parameter the rung singlet-phase of both models $\mathcal{H}_{\shortparallel}$ and $\mathcal{H}_{\bowtie}$ for all relevant $\lambda$ is not topologically protected but trivially connected to the product state of rung singlets \cite{Pollmann2012}.

%%%%%%%%%%%%%%%%%%%%%%%%%%%%%%%%%%%%%%%%%%%%%%%%%%%%%%%%%%%%%%%%%%%%%%%%%%%%%%%%%%%%%%%%%%%% 
%Checkout:
%
%Entanglement and fluctuations in the XXZ model with power-law interactions (Frérot)
%https://journals.aps.org/prb/abstract/10.1103/PhysRevB.95.245111
%
%Criticality and phase diagram of quantum long-range O(N) models (Defenu)
%https://journals.aps.org/prb/abstract/10.1103/PhysRevB.96.104432
%
%Deconfined quantum criticality in spin-1/2 chains with long-range interactions (YangSandvik2020)
%https://arxiv.org/abs/2001.02821
%
%Locality of gapped ground states in systems with power-law decaying interaction:
%https://arxiv.org/pdf/2208.13057.pdf
%
%Generalized Higgs mechanism in long-range interacting quantum systems
%https://arxiv.org/abs/2208.10487
%
%Symmetry-protected topological phases in spin ladders with two-body interactions
%https://journals.aps.org/prb/abstract/10.1103/PhysRevB.86.195122
%%%%%%%%%%%%%%%%%%%%%%%%%%%%%%%%%%%%%%%%%%%%%%%%%%%%%%%%%%%%%%%%%%%%%%%%%%%%%%%%%%%%%%%%%%%% 

% Conclusion
%%%%%%%%%%%%%%%%%%%%%%%%%%%%%%%%%%%%%%%%%%%%%%%%%%%%%%%%%%%%%%%%%%%%%%%%%%%%%%%%%%%%%%%%%%%% 
{\color{blue}\emph{Conclusions.-}}    
We investigated the quantum-critical behavior of unfrustrated two-leg quantum spin ladders with long-range Heisenberg interactions by applying and extending the pCUT method using white graphs and classical Monte Carlo integration. Calculating the ground-state energy, the one-triplon gap, and the one-triplon spectral weight allows us to extract the full set of critical exponents as a function of the decay exponent by appropriate extrapolation techniques. A non-trivial regime of continuously varying critical exponents as well as long-range mean-field behavior is observed similar to the unfrustrated LRTFIM implying the absence of deconfined criticality. Let us note that our approach can be naturally extended to two-dimensional models with long-range interactions like Heisenberg bilayers offering a completely unexplored playground for future investigations.

{\color{blue}\emph{Acknowledgments.-}}
PA and KPS gratefully acknowledge the support by the Deutsche Forschungsgemeinschaft (DFG, German Research
Foundation) -- Project-ID 429529648—TRR 306 QuCoLiMa (``Quantum Cooperativity of Light and Matter'') as well as the Munich Quantum Valley, which is supported by the Bavarian state government with funds from the Hightech Agenda Bayern Plus and the scientific support and HPC resources provided by the Erlangen National High Performance Computing Center (NHR@FAU) of the Friedrich-Alexander-Universität Erlangen-Nürnberg (FAU). The hardware is funded by the DFG.

%{\color{blue}\emph{Data Availability.-}}
%The data is available on Zenodo \patricknew{XXX}.

%{\bf Supplementary information:} The online version contains supplementary material
%available at: http://url.
%\bibliography{bibliography.bib}

\begin{thebibliography}{91}%
\makeatletter
\providecommand \@ifxundefined [1]{%
 \@ifx{#1\undefined}
}%
\providecommand \@ifnum [1]{%
 \ifnum #1\expandafter \@firstoftwo
 \else \expandafter \@secondoftwo
 \fi
}%
\providecommand \@ifx [1]{%
 \ifx #1\expandafter \@firstoftwo
 \else \expandafter \@secondoftwo
 \fi
}%
\providecommand \natexlab [1]{#1}%
\providecommand \enquote  [1]{``#1''}%
\providecommand \bibnamefont  [1]{#1}%
\providecommand \bibfnamefont [1]{#1}%
\providecommand \citenamefont [1]{#1}%
\providecommand \href@noop [0]{\@secondoftwo}%
\providecommand \href [0]{\begingroup \@sanitize@url \@href}%
\providecommand \@href[1]{\@@startlink{#1}\@@href}%
\providecommand \@@href[1]{\endgroup#1\@@endlink}%
\providecommand \@sanitize@url [0]{\catcode `\\12\catcode `\$12\catcode
  `\&12\catcode `\#12\catcode `\^12\catcode `\_12\catcode `\%12\relax}%
\providecommand \@@startlink[1]{}%
\providecommand \@@endlink[0]{}%
\providecommand \url  [0]{\begingroup\@sanitize@url \@url }%
\providecommand \@url [1]{\endgroup\@href {#1}{\urlprefix }}%
\providecommand \urlprefix  [0]{URL }%
\providecommand \Eprint [0]{\href }%
\providecommand \doibase [0]{http://dx.doi.org/}%
\providecommand \selectlanguage [0]{\@gobble}%
\providecommand \bibinfo  [0]{\@secondoftwo}%
\providecommand \bibfield  [0]{\@secondoftwo}%
\providecommand \translation [1]{[#1]}%
\providecommand \BibitemOpen [0]{}%
\providecommand \bibitemStop [0]{}%
\providecommand \bibitemNoStop [0]{.\EOS\space}%
\providecommand \EOS [0]{\spacefactor3000\relax}%
\providecommand \BibitemShut  [1]{\csname bibitem#1\endcsname}%
\let\auto@bib@innerbib\@empty
%</preamble>
\bibitem [{\citenamefont {Bitko}\ \emph {et~al.}(1996)\citenamefont {Bitko},
  \citenamefont {Rosenbaum},\ and\ \citenamefont {Aeppli}}]{Bitko1996}%
  \BibitemOpen
  \bibfield  {author} {\bibinfo {author} {\bibfnamefont {D.}~\bibnamefont
  {Bitko}}, \bibinfo {author} {\bibfnamefont {T.~F.}\ \bibnamefont
  {Rosenbaum}}, \ and\ \bibinfo {author} {\bibfnamefont {G.}~\bibnamefont
  {Aeppli}},\ }\href {\doibase 10.1103/PhysRevLett.77.940} {\bibfield
  {journal} {\bibinfo  {journal} {Phys. Rev. Lett.}\ }\textbf {\bibinfo
  {volume} {77}},\ \bibinfo {pages} {940} (\bibinfo {year} {1996})}\BibitemShut
  {NoStop}%
\bibitem [{\citenamefont {Chakraborty}\ \emph {et~al.}(2004)\citenamefont
  {Chakraborty}, \citenamefont {Henelius}, \citenamefont {Kj\o{}nsberg},
  \citenamefont {Sandvik},\ and\ \citenamefont {Girvin}}]{Chakraborty2004}%
  \BibitemOpen
  \bibfield  {author} {\bibinfo {author} {\bibfnamefont {P.~B.}\ \bibnamefont
  {Chakraborty}}, \bibinfo {author} {\bibfnamefont {P.}~\bibnamefont
  {Henelius}}, \bibinfo {author} {\bibfnamefont {H.}~\bibnamefont
  {Kj\o{}nsberg}}, \bibinfo {author} {\bibfnamefont {A.~W.}\ \bibnamefont
  {Sandvik}}, \ and\ \bibinfo {author} {\bibfnamefont {S.~M.}\ \bibnamefont
  {Girvin}},\ }\href {\doibase 10.1103/PhysRevB.70.144411} {\bibfield
  {journal} {\bibinfo  {journal} {Phys. Rev. B}\ }\textbf {\bibinfo {volume}
  {70}},\ \bibinfo {pages} {144411} (\bibinfo {year} {2004})}\BibitemShut
  {NoStop}%
\bibitem [{\citenamefont {Bramwell}\ and\ \citenamefont
  {Gingras}(2001)}]{Bramwell2001}%
  \BibitemOpen
  \bibfield  {author} {\bibinfo {author} {\bibfnamefont {S.~T.}\ \bibnamefont
  {Bramwell}}\ and\ \bibinfo {author} {\bibfnamefont {M.~J.~P.}\ \bibnamefont
  {Gingras}},\ }\href {\doibase 10.1126/science.1064761} {\bibfield  {journal}
  {\bibinfo  {journal} {Science}\ }\textbf {\bibinfo {volume} {294}},\ \bibinfo
  {pages} {1495} (\bibinfo {year} {2001})}\BibitemShut {NoStop}%
\bibitem [{\citenamefont {Castelnovo}\ \emph {et~al.}(2008)\citenamefont
  {Castelnovo}, \citenamefont {Moessner},\ and\ \citenamefont
  {Sondhi}}]{Castelnovo2008}%
  \BibitemOpen
  \bibfield  {author} {\bibinfo {author} {\bibfnamefont {C.}~\bibnamefont
  {Castelnovo}}, \bibinfo {author} {\bibfnamefont {R.}~\bibnamefont
  {Moessner}}, \ and\ \bibinfo {author} {\bibfnamefont {S.~L.}\ \bibnamefont
  {Sondhi}},\ }\href {\doibase 10.1038/nature06433} {\bibfield  {journal}
  {\bibinfo  {journal} {Nature}\ }\textbf {\bibinfo {volume} {452}},\ \bibinfo
  {pages} {43} (\bibinfo {year} {2008})}\BibitemShut {NoStop}%
\bibitem [{\citenamefont {Islam}\ \emph {et~al.}(2011)\citenamefont {Islam},
  \citenamefont {Edwards}, \citenamefont {Kim}, \citenamefont {Korenblit},
  \citenamefont {Noh}, \citenamefont {Carmichael}, \citenamefont {Lin},
  \citenamefont {Duan}, \citenamefont {Joseph~Wang}, \citenamefont
  {Freericks},\ and\ \citenamefont {Monroe}}]{Islam2011}%
  \BibitemOpen
  \bibfield  {author} {\bibinfo {author} {\bibfnamefont {R.}~\bibnamefont
  {Islam}}, \bibinfo {author} {\bibfnamefont {E.~E.}\ \bibnamefont {Edwards}},
  \bibinfo {author} {\bibfnamefont {K.}~\bibnamefont {Kim}}, \bibinfo {author}
  {\bibfnamefont {S.}~\bibnamefont {Korenblit}}, \bibinfo {author}
  {\bibfnamefont {C.}~\bibnamefont {Noh}}, \bibinfo {author} {\bibfnamefont
  {H.}~\bibnamefont {Carmichael}}, \bibinfo {author} {\bibfnamefont {G.-D.}\
  \bibnamefont {Lin}}, \bibinfo {author} {\bibfnamefont {L.-M.}\ \bibnamefont
  {Duan}}, \bibinfo {author} {\bibfnamefont {C.-C.}\ \bibnamefont
  {Joseph~Wang}}, \bibinfo {author} {\bibfnamefont {J.~K.}\ \bibnamefont
  {Freericks}}, \ and\ \bibinfo {author} {\bibfnamefont {C.}~\bibnamefont
  {Monroe}},\ }\href {\doibase 10.1038/ncomms1374} {\bibfield  {journal}
  {\bibinfo  {journal} {Nature Communications}\ }\textbf {\bibinfo {volume}
  {2}},\ \bibinfo {pages} {377} (\bibinfo {year} {2011})}\BibitemShut {NoStop}%
\bibitem [{\citenamefont {Britton}\ \emph {et~al.}(2012)\citenamefont
  {Britton}, \citenamefont {Sawyer}, \citenamefont {Keith}, \citenamefont
  {Wang}, \citenamefont {Freericks}, \citenamefont {Uys}, \citenamefont
  {Biercuk},\ and\ \citenamefont {Bollinger}}]{Britton2012}%
  \BibitemOpen
  \bibfield  {author} {\bibinfo {author} {\bibfnamefont {J.~W.}\ \bibnamefont
  {Britton}}, \bibinfo {author} {\bibfnamefont {B.~C.}\ \bibnamefont {Sawyer}},
  \bibinfo {author} {\bibfnamefont {A.~C.}\ \bibnamefont {Keith}}, \bibinfo
  {author} {\bibfnamefont {C.-C.~J.}\ \bibnamefont {Wang}}, \bibinfo {author}
  {\bibfnamefont {J.~K.}\ \bibnamefont {Freericks}}, \bibinfo {author}
  {\bibfnamefont {H.}~\bibnamefont {Uys}}, \bibinfo {author} {\bibfnamefont
  {M.~J.}\ \bibnamefont {Biercuk}}, \ and\ \bibinfo {author} {\bibfnamefont
  {J.~J.}\ \bibnamefont {Bollinger}},\ }\href {\doibase 10.1038/nature10981}
  {\bibfield  {journal} {\bibinfo  {journal} {Nature}\ }\textbf {\bibinfo
  {volume} {484}},\ \bibinfo {pages} {489} (\bibinfo {year}
  {2012})}\BibitemShut {NoStop}%
\bibitem [{\citenamefont {Islam}\ \emph {et~al.}(2013)\citenamefont {Islam},
  \citenamefont {Senko}, \citenamefont {Campbell}, \citenamefont {Korenblit},
  \citenamefont {Smith}, \citenamefont {Lee}, \citenamefont {Edwards},
  \citenamefont {Wang}, \citenamefont {Freericks},\ and\ \citenamefont
  {Monroe}}]{Islam2013}%
  \BibitemOpen
  \bibfield  {author} {\bibinfo {author} {\bibfnamefont {R.}~\bibnamefont
  {Islam}}, \bibinfo {author} {\bibfnamefont {C.}~\bibnamefont {Senko}},
  \bibinfo {author} {\bibfnamefont {W.~C.}\ \bibnamefont {Campbell}}, \bibinfo
  {author} {\bibfnamefont {S.}~\bibnamefont {Korenblit}}, \bibinfo {author}
  {\bibfnamefont {J.}~\bibnamefont {Smith}}, \bibinfo {author} {\bibfnamefont
  {A.}~\bibnamefont {Lee}}, \bibinfo {author} {\bibfnamefont {E.~E.}\
  \bibnamefont {Edwards}}, \bibinfo {author} {\bibfnamefont {C.-C.~J.}\
  \bibnamefont {Wang}}, \bibinfo {author} {\bibfnamefont {J.~K.}\ \bibnamefont
  {Freericks}}, \ and\ \bibinfo {author} {\bibfnamefont {C.}~\bibnamefont
  {Monroe}},\ }\href {\doibase 10.1126/science.1232296} {\bibfield  {journal}
  {\bibinfo  {journal} {Science}\ }\textbf {\bibinfo {volume} {340}},\ \bibinfo
  {pages} {583} (\bibinfo {year} {2013})}\BibitemShut {NoStop}%
\bibitem [{\citenamefont {Jurcevic}\ \emph {et~al.}(2014)\citenamefont
  {Jurcevic}, \citenamefont {Lanyon}, \citenamefont {Hauke}, \citenamefont
  {Hempel}, \citenamefont {Zoller}, \citenamefont {Blatt},\ and\ \citenamefont
  {Roos}}]{Jurcevic2014}%
  \BibitemOpen
  \bibfield  {author} {\bibinfo {author} {\bibfnamefont {P.}~\bibnamefont
  {Jurcevic}}, \bibinfo {author} {\bibfnamefont {B.~P.}\ \bibnamefont
  {Lanyon}}, \bibinfo {author} {\bibfnamefont {P.}~\bibnamefont {Hauke}},
  \bibinfo {author} {\bibfnamefont {C.}~\bibnamefont {Hempel}}, \bibinfo
  {author} {\bibfnamefont {P.}~\bibnamefont {Zoller}}, \bibinfo {author}
  {\bibfnamefont {R.}~\bibnamefont {Blatt}}, \ and\ \bibinfo {author}
  {\bibfnamefont {C.~F.}\ \bibnamefont {Roos}},\ }\href {\doibase
  10.1038/nature13461} {\bibfield  {journal} {\bibinfo  {journal} {Nature}\
  }\textbf {\bibinfo {volume} {511}},\ \bibinfo {pages} {202} (\bibinfo {year}
  {2014})}\BibitemShut {NoStop}%
\bibitem [{\citenamefont {Richerme}\ \emph {et~al.}(2014)\citenamefont
  {Richerme}, \citenamefont {Gong}, \citenamefont {Lee}, \citenamefont {Senko},
  \citenamefont {Smith}, \citenamefont {Foss-Feig}, \citenamefont {Michalakis},
  \citenamefont {Gorshkov},\ and\ \citenamefont {Monroe}}]{Richerme2014}%
  \BibitemOpen
  \bibfield  {author} {\bibinfo {author} {\bibfnamefont {P.}~\bibnamefont
  {Richerme}}, \bibinfo {author} {\bibfnamefont {Z.-X.}\ \bibnamefont {Gong}},
  \bibinfo {author} {\bibfnamefont {A.}~\bibnamefont {Lee}}, \bibinfo {author}
  {\bibfnamefont {C.}~\bibnamefont {Senko}}, \bibinfo {author} {\bibfnamefont
  {J.}~\bibnamefont {Smith}}, \bibinfo {author} {\bibfnamefont
  {M.}~\bibnamefont {Foss-Feig}}, \bibinfo {author} {\bibfnamefont
  {S.}~\bibnamefont {Michalakis}}, \bibinfo {author} {\bibfnamefont {A.~V.}\
  \bibnamefont {Gorshkov}}, \ and\ \bibinfo {author} {\bibfnamefont
  {C.}~\bibnamefont {Monroe}},\ }\href {\doibase 10.1038/nature13450}
  {\bibfield  {journal} {\bibinfo  {journal} {Nature}\ }\textbf {\bibinfo
  {volume} {511}},\ \bibinfo {pages} {198} (\bibinfo {year}
  {2014})}\BibitemShut {NoStop}%
\bibitem [{\citenamefont {Mielenz}\ \emph {et~al.}(2016)\citenamefont
  {Mielenz}, \citenamefont {Kalis}, \citenamefont {Wittemer}, \citenamefont
  {Hakelberg}, \citenamefont {Warring}, \citenamefont {Schmied}, \citenamefont
  {Blain}, \citenamefont {Maunz}, \citenamefont {Moehring}, \citenamefont
  {Leibfried},\ and\ \citenamefont {Schaetz}}]{Mielenz2016}%
  \BibitemOpen
  \bibfield  {author} {\bibinfo {author} {\bibfnamefont {M.}~\bibnamefont
  {Mielenz}}, \bibinfo {author} {\bibfnamefont {H.}~\bibnamefont {Kalis}},
  \bibinfo {author} {\bibfnamefont {M.}~\bibnamefont {Wittemer}}, \bibinfo
  {author} {\bibfnamefont {F.}~\bibnamefont {Hakelberg}}, \bibinfo {author}
  {\bibfnamefont {U.}~\bibnamefont {Warring}}, \bibinfo {author} {\bibfnamefont
  {R.}~\bibnamefont {Schmied}}, \bibinfo {author} {\bibfnamefont
  {M.}~\bibnamefont {Blain}}, \bibinfo {author} {\bibfnamefont
  {P.}~\bibnamefont {Maunz}}, \bibinfo {author} {\bibfnamefont {D.~L.}\
  \bibnamefont {Moehring}}, \bibinfo {author} {\bibfnamefont {D.}~\bibnamefont
  {Leibfried}}, \ and\ \bibinfo {author} {\bibfnamefont {T.}~\bibnamefont
  {Schaetz}},\ }\href {\doibase 10.1038/ncomms11839} {\bibfield  {journal}
  {\bibinfo  {journal} {Nature Communications}\ }\textbf {\bibinfo {volume}
  {7}},\ \bibinfo {pages} {ncomms11839} (\bibinfo {year} {2016})}\BibitemShut
  {NoStop}%
\bibitem [{\citenamefont {Bohnet}\ \emph {et~al.}(2016)\citenamefont {Bohnet},
  \citenamefont {Sawyer}, \citenamefont {Britton}, \citenamefont {Wall},
  \citenamefont {Rey}, \citenamefont {Foss-Feig},\ and\ \citenamefont
  {Bollinger}}]{Bohnet2016}%
  \BibitemOpen
  \bibfield  {author} {\bibinfo {author} {\bibfnamefont {J.~G.}\ \bibnamefont
  {Bohnet}}, \bibinfo {author} {\bibfnamefont {B.~C.}\ \bibnamefont {Sawyer}},
  \bibinfo {author} {\bibfnamefont {J.~W.}\ \bibnamefont {Britton}}, \bibinfo
  {author} {\bibfnamefont {M.~L.}\ \bibnamefont {Wall}}, \bibinfo {author}
  {\bibfnamefont {A.~M.}\ \bibnamefont {Rey}}, \bibinfo {author} {\bibfnamefont
  {M.}~\bibnamefont {Foss-Feig}}, \ and\ \bibinfo {author} {\bibfnamefont
  {J.~J.}\ \bibnamefont {Bollinger}},\ }\href {\doibase
  10.1126/science.aad9958} {\bibfield  {journal} {\bibinfo  {journal}
  {Science}\ }\textbf {\bibinfo {volume} {352}},\ \bibinfo {pages} {1297}
  (\bibinfo {year} {2016})}\BibitemShut {NoStop}%
\bibitem [{\citenamefont {Jurcevic}\ \emph {et~al.}(2017)\citenamefont
  {Jurcevic}, \citenamefont {Shen}, \citenamefont {Hauke}, \citenamefont
  {Maier}, \citenamefont {Brydges}, \citenamefont {Hempel}, \citenamefont
  {Lanyon}, \citenamefont {Heyl}, \citenamefont {Blatt},\ and\ \citenamefont
  {Roos}}]{Jurcevic2017}%
  \BibitemOpen
  \bibfield  {author} {\bibinfo {author} {\bibfnamefont {P.}~\bibnamefont
  {Jurcevic}}, \bibinfo {author} {\bibfnamefont {H.}~\bibnamefont {Shen}},
  \bibinfo {author} {\bibfnamefont {P.}~\bibnamefont {Hauke}}, \bibinfo
  {author} {\bibfnamefont {C.}~\bibnamefont {Maier}}, \bibinfo {author}
  {\bibfnamefont {T.}~\bibnamefont {Brydges}}, \bibinfo {author} {\bibfnamefont
  {C.}~\bibnamefont {Hempel}}, \bibinfo {author} {\bibfnamefont {B.~P.}\
  \bibnamefont {Lanyon}}, \bibinfo {author} {\bibfnamefont {M.}~\bibnamefont
  {Heyl}}, \bibinfo {author} {\bibfnamefont {R.}~\bibnamefont {Blatt}}, \ and\
  \bibinfo {author} {\bibfnamefont {C.~F.}\ \bibnamefont {Roos}},\ }\href
  {\doibase 10.1103/PhysRevLett.119.080501} {\bibfield  {journal} {\bibinfo
  {journal} {Phys. Rev. Lett.}\ }\textbf {\bibinfo {volume} {119}},\ \bibinfo
  {pages} {080501} (\bibinfo {year} {2017})}\BibitemShut {NoStop}%
\bibitem [{\citenamefont {Zhang}\ \emph {et~al.}(2017)\citenamefont {Zhang},
  \citenamefont {Pagano}, \citenamefont {Hess}, \citenamefont {Kyprianidis},
  \citenamefont {Becker}, \citenamefont {Kaplan}, \citenamefont {Gorshkov},
  \citenamefont {Gong},\ and\ \citenamefont {Monroe}}]{Zhang2017}%
  \BibitemOpen
  \bibfield  {author} {\bibinfo {author} {\bibfnamefont {J.}~\bibnamefont
  {Zhang}}, \bibinfo {author} {\bibfnamefont {G.}~\bibnamefont {Pagano}},
  \bibinfo {author} {\bibfnamefont {P.~W.}\ \bibnamefont {Hess}}, \bibinfo
  {author} {\bibfnamefont {A.}~\bibnamefont {Kyprianidis}}, \bibinfo {author}
  {\bibfnamefont {P.}~\bibnamefont {Becker}}, \bibinfo {author} {\bibfnamefont
  {H.}~\bibnamefont {Kaplan}}, \bibinfo {author} {\bibfnamefont {A.~V.}\
  \bibnamefont {Gorshkov}}, \bibinfo {author} {\bibfnamefont {Z.-X.}\
  \bibnamefont {Gong}}, \ and\ \bibinfo {author} {\bibfnamefont
  {C.}~\bibnamefont {Monroe}},\ }\href {\doibase 10.1038/nature24654}
  {\bibfield  {journal} {\bibinfo  {journal} {Nature}\ }\textbf {\bibinfo
  {volume} {551}},\ \bibinfo {pages} {601} (\bibinfo {year}
  {2017})}\BibitemShut {NoStop}%
\bibitem [{\citenamefont {\ifmmode \check{Z}\else
  \v{Z}\fi{}unkovi\ifmmode~\check{c}\else \v{c}\fi{}}\ \emph
  {et~al.}(2018)\citenamefont {\ifmmode \check{Z}\else
  \v{Z}\fi{}unkovi\ifmmode~\check{c}\else \v{c}\fi{}}, \citenamefont {Heyl},
  \citenamefont {Knap},\ and\ \citenamefont {Silva}}]{Zunkovic2018}%
  \BibitemOpen
  \bibfield  {author} {\bibinfo {author} {\bibfnamefont {B.}~\bibnamefont
  {\ifmmode \check{Z}\else \v{Z}\fi{}unkovi\ifmmode~\check{c}\else
  \v{c}\fi{}}}, \bibinfo {author} {\bibfnamefont {M.}~\bibnamefont {Heyl}},
  \bibinfo {author} {\bibfnamefont {M.}~\bibnamefont {Knap}}, \ and\ \bibinfo
  {author} {\bibfnamefont {A.}~\bibnamefont {Silva}},\ }\href {\doibase
  10.1103/PhysRevLett.120.130601} {\bibfield  {journal} {\bibinfo  {journal}
  {Phys. Rev. Lett.}\ }\textbf {\bibinfo {volume} {120}},\ \bibinfo {pages}
  {130601} (\bibinfo {year} {2018})}\BibitemShut {NoStop}%
\bibitem [{\citenamefont {Hempel}\ \emph {et~al.}(2018)\citenamefont {Hempel},
  \citenamefont {Maier}, \citenamefont {Romero}, \citenamefont {McClean},
  \citenamefont {Monz}, \citenamefont {Shen}, \citenamefont {Jurcevic},
  \citenamefont {Lanyon}, \citenamefont {Love}, \citenamefont {Babbush},
  \citenamefont {Aspuru-Guzik}, \citenamefont {Blatt},\ and\ \citenamefont
  {Roos}}]{Hempel2018}%
  \BibitemOpen
  \bibfield  {author} {\bibinfo {author} {\bibfnamefont {C.}~\bibnamefont
  {Hempel}}, \bibinfo {author} {\bibfnamefont {C.}~\bibnamefont {Maier}},
  \bibinfo {author} {\bibfnamefont {J.}~\bibnamefont {Romero}}, \bibinfo
  {author} {\bibfnamefont {J.}~\bibnamefont {McClean}}, \bibinfo {author}
  {\bibfnamefont {T.}~\bibnamefont {Monz}}, \bibinfo {author} {\bibfnamefont
  {H.}~\bibnamefont {Shen}}, \bibinfo {author} {\bibfnamefont {P.}~\bibnamefont
  {Jurcevic}}, \bibinfo {author} {\bibfnamefont {B.~P.}\ \bibnamefont
  {Lanyon}}, \bibinfo {author} {\bibfnamefont {P.}~\bibnamefont {Love}},
  \bibinfo {author} {\bibfnamefont {R.}~\bibnamefont {Babbush}}, \bibinfo
  {author} {\bibfnamefont {A.}~\bibnamefont {Aspuru-Guzik}}, \bibinfo {author}
  {\bibfnamefont {R.}~\bibnamefont {Blatt}}, \ and\ \bibinfo {author}
  {\bibfnamefont {C.~F.}\ \bibnamefont {Roos}},\ }\href {\doibase
  10.1103/PhysRevX.8.031022} {\bibfield  {journal} {\bibinfo  {journal} {Phys.
  Rev. X}\ }\textbf {\bibinfo {volume} {8}},\ \bibinfo {pages} {031022}
  (\bibinfo {year} {2018})}\BibitemShut {NoStop}%
\bibitem [{\citenamefont {Joshi}\ \emph {et~al.}(2022)\citenamefont {Joshi},
  \citenamefont {Kranzl}, \citenamefont {Schuckert}, \citenamefont {Lovas},
  \citenamefont {Maier}, \citenamefont {Blatt}, \citenamefont {Knap},\ and\
  \citenamefont {Roos}}]{Joshi2022}%
  \BibitemOpen
  \bibfield  {author} {\bibinfo {author} {\bibfnamefont {M.~K.}\ \bibnamefont
  {Joshi}}, \bibinfo {author} {\bibfnamefont {F.}~\bibnamefont {Kranzl}},
  \bibinfo {author} {\bibfnamefont {A.}~\bibnamefont {Schuckert}}, \bibinfo
  {author} {\bibfnamefont {I.}~\bibnamefont {Lovas}}, \bibinfo {author}
  {\bibfnamefont {C.}~\bibnamefont {Maier}}, \bibinfo {author} {\bibfnamefont
  {R.}~\bibnamefont {Blatt}}, \bibinfo {author} {\bibfnamefont
  {M.}~\bibnamefont {Knap}}, \ and\ \bibinfo {author} {\bibfnamefont {C.~F.}\
  \bibnamefont {Roos}},\ }\href {\doibase 10.1126/science.abk2400} {\bibfield
  {journal} {\bibinfo  {journal} {Science}\ }\textbf {\bibinfo {volume}
  {376}},\ \bibinfo {pages} {720} (\bibinfo {year} {2022})}\BibitemShut
  {NoStop}%
\bibitem [{\citenamefont {Weimer}\ \emph {et~al.}(2010)\citenamefont {Weimer},
  \citenamefont {M{\"u}ller}, \citenamefont {Lesanovsky}, \citenamefont
  {Zoller},\ and\ \citenamefont {B{\"u}chler}}]{Weimer2010}%
  \BibitemOpen
  \bibfield  {author} {\bibinfo {author} {\bibfnamefont {H.}~\bibnamefont
  {Weimer}}, \bibinfo {author} {\bibfnamefont {M.}~\bibnamefont {M{\"u}ller}},
  \bibinfo {author} {\bibfnamefont {I.}~\bibnamefont {Lesanovsky}}, \bibinfo
  {author} {\bibfnamefont {P.}~\bibnamefont {Zoller}}, \ and\ \bibinfo {author}
  {\bibfnamefont {H.~P.}\ \bibnamefont {B{\"u}chler}},\ }\href {\doibase
  10.1038/nphys1614} {\bibfield  {journal} {\bibinfo  {journal} {Nature
  Physics}\ }\textbf {\bibinfo {volume} {6}},\ \bibinfo {pages} {382} (\bibinfo
  {year} {2010})}\BibitemShut {NoStop}%
\bibitem [{\citenamefont {Xia}\ \emph {et~al.}(2015)\citenamefont {Xia},
  \citenamefont {Lichtman}, \citenamefont {Maller}, \citenamefont {Carr},
  \citenamefont {Piotrowicz}, \citenamefont {Isenhower},\ and\ \citenamefont
  {Saffman}}]{Xia2015}%
  \BibitemOpen
  \bibfield  {author} {\bibinfo {author} {\bibfnamefont {T.}~\bibnamefont
  {Xia}}, \bibinfo {author} {\bibfnamefont {M.}~\bibnamefont {Lichtman}},
  \bibinfo {author} {\bibfnamefont {K.}~\bibnamefont {Maller}}, \bibinfo
  {author} {\bibfnamefont {A.~W.}\ \bibnamefont {Carr}}, \bibinfo {author}
  {\bibfnamefont {M.~J.}\ \bibnamefont {Piotrowicz}}, \bibinfo {author}
  {\bibfnamefont {L.}~\bibnamefont {Isenhower}}, \ and\ \bibinfo {author}
  {\bibfnamefont {M.}~\bibnamefont {Saffman}},\ }\href {\doibase
  10.1103/PhysRevLett.114.100503} {\bibfield  {journal} {\bibinfo  {journal}
  {Phys. Rev. Lett.}\ }\textbf {\bibinfo {volume} {114}},\ \bibinfo {pages}
  {100503} (\bibinfo {year} {2015})}\BibitemShut {NoStop}%
\bibitem [{\citenamefont {Labuhn}\ \emph {et~al.}(2016)\citenamefont {Labuhn},
  \citenamefont {Barredo}, \citenamefont {Ravets}, \citenamefont
  {de~L{\'e}s{\'e}leuc}, \citenamefont {Macr{\`i}}, \citenamefont {Lahaye},\
  and\ \citenamefont {Browaeys}}]{Labuhn2016}%
  \BibitemOpen
  \bibfield  {author} {\bibinfo {author} {\bibfnamefont {H.}~\bibnamefont
  {Labuhn}}, \bibinfo {author} {\bibfnamefont {D.}~\bibnamefont {Barredo}},
  \bibinfo {author} {\bibfnamefont {S.}~\bibnamefont {Ravets}}, \bibinfo
  {author} {\bibfnamefont {S.}~\bibnamefont {de~L{\'e}s{\'e}leuc}}, \bibinfo
  {author} {\bibfnamefont {T.}~\bibnamefont {Macr{\`i}}}, \bibinfo {author}
  {\bibfnamefont {T.}~\bibnamefont {Lahaye}}, \ and\ \bibinfo {author}
  {\bibfnamefont {A.}~\bibnamefont {Browaeys}},\ }\href {\doibase
  10.1038/nature18274} {\bibfield  {journal} {\bibinfo  {journal} {Nature}\
  }\textbf {\bibinfo {volume} {534}},\ \bibinfo {pages} {667} (\bibinfo {year}
  {2016})}\BibitemShut {NoStop}%
\bibitem [{\citenamefont {Wang}\ \emph {et~al.}(2016)\citenamefont {Wang},
  \citenamefont {Kumar}, \citenamefont {Wu},\ and\ \citenamefont
  {Weiss}}]{Wang2016}%
  \BibitemOpen
  \bibfield  {author} {\bibinfo {author} {\bibfnamefont {Y.}~\bibnamefont
  {Wang}}, \bibinfo {author} {\bibfnamefont {A.}~\bibnamefont {Kumar}},
  \bibinfo {author} {\bibfnamefont {T.-Y.}\ \bibnamefont {Wu}}, \ and\ \bibinfo
  {author} {\bibfnamefont {D.~S.}\ \bibnamefont {Weiss}},\ }\href {\doibase
  10.1126/science.aaf2581} {\bibfield  {journal} {\bibinfo  {journal}
  {Science}\ }\textbf {\bibinfo {volume} {352}},\ \bibinfo {pages} {1562}
  (\bibinfo {year} {2016})}\BibitemShut {NoStop}%
\bibitem [{\citenamefont {Schauss}(2018)}]{Schauss2018}%
  \BibitemOpen
  \bibfield  {author} {\bibinfo {author} {\bibfnamefont {P.}~\bibnamefont
  {Schauss}},\ }\href {\doibase 10.1088/2058-9565/aa9c59} {\bibfield  {journal}
  {\bibinfo  {journal} {Quantum Science and Technology}\ }\textbf {\bibinfo
  {volume} {3}},\ \bibinfo {pages} {023001} (\bibinfo {year}
  {2018})}\BibitemShut {NoStop}%
\bibitem [{\citenamefont {de~Léséleuc}\ \emph {et~al.}(2019)\citenamefont
  {de~Léséleuc}, \citenamefont {Lienhard}, \citenamefont {Scholl},
  \citenamefont {Barredo}, \citenamefont {Weber}, \citenamefont {Lang},
  \citenamefont {Büchler}, \citenamefont {Lahaye},\ and\ \citenamefont
  {Browaeys}}]{Leseleuc2019}%
  \BibitemOpen
  \bibfield  {author} {\bibinfo {author} {\bibfnamefont {S.}~\bibnamefont
  {de~Léséleuc}}, \bibinfo {author} {\bibfnamefont {V.}~\bibnamefont
  {Lienhard}}, \bibinfo {author} {\bibfnamefont {P.}~\bibnamefont {Scholl}},
  \bibinfo {author} {\bibfnamefont {D.}~\bibnamefont {Barredo}}, \bibinfo
  {author} {\bibfnamefont {S.}~\bibnamefont {Weber}}, \bibinfo {author}
  {\bibfnamefont {N.}~\bibnamefont {Lang}}, \bibinfo {author} {\bibfnamefont
  {H.~P.}\ \bibnamefont {Büchler}}, \bibinfo {author} {\bibfnamefont
  {T.}~\bibnamefont {Lahaye}}, \ and\ \bibinfo {author} {\bibfnamefont
  {A.}~\bibnamefont {Browaeys}},\ }\href {\doibase 10.1126/science.aav9105}
  {\bibfield  {journal} {\bibinfo  {journal} {Science}\ }\textbf {\bibinfo
  {volume} {365}},\ \bibinfo {pages} {775} (\bibinfo {year}
  {2019})}\BibitemShut {NoStop}%
\bibitem [{\citenamefont {Levine}\ \emph {et~al.}(2019)\citenamefont {Levine},
  \citenamefont {Keesling}, \citenamefont {Semeghini}, \citenamefont {Omran},
  \citenamefont {Wang}, \citenamefont {Ebadi}, \citenamefont {Bernien},
  \citenamefont {Greiner}, \citenamefont {Vuleti\ifmmode~\acute{c}\else
  \'{c}\fi{}}, \citenamefont {Pichler},\ and\ \citenamefont
  {Lukin}}]{Levine2019}%
  \BibitemOpen
  \bibfield  {author} {\bibinfo {author} {\bibfnamefont {H.}~\bibnamefont
  {Levine}}, \bibinfo {author} {\bibfnamefont {A.}~\bibnamefont {Keesling}},
  \bibinfo {author} {\bibfnamefont {G.}~\bibnamefont {Semeghini}}, \bibinfo
  {author} {\bibfnamefont {A.}~\bibnamefont {Omran}}, \bibinfo {author}
  {\bibfnamefont {T.~T.}\ \bibnamefont {Wang}}, \bibinfo {author}
  {\bibfnamefont {S.}~\bibnamefont {Ebadi}}, \bibinfo {author} {\bibfnamefont
  {H.}~\bibnamefont {Bernien}}, \bibinfo {author} {\bibfnamefont
  {M.}~\bibnamefont {Greiner}}, \bibinfo {author} {\bibfnamefont
  {V.}~\bibnamefont {Vuleti\ifmmode~\acute{c}\else \'{c}\fi{}}}, \bibinfo
  {author} {\bibfnamefont {H.}~\bibnamefont {Pichler}}, \ and\ \bibinfo
  {author} {\bibfnamefont {M.~D.}\ \bibnamefont {Lukin}},\ }\href {\doibase
  10.1103/PhysRevLett.123.170503} {\bibfield  {journal} {\bibinfo  {journal}
  {Phys. Rev. Lett.}\ }\textbf {\bibinfo {volume} {123}},\ \bibinfo {pages}
  {170503} (\bibinfo {year} {2019})}\BibitemShut {NoStop}%
\bibitem [{\citenamefont {Wu}\ \emph {et~al.}(2019)\citenamefont {Wu},
  \citenamefont {Kumar}, \citenamefont {Giraldo},\ and\ \citenamefont
  {Weiss}}]{Wu2019}%
  \BibitemOpen
  \bibfield  {author} {\bibinfo {author} {\bibfnamefont {T.-Y.}\ \bibnamefont
  {Wu}}, \bibinfo {author} {\bibfnamefont {A.}~\bibnamefont {Kumar}}, \bibinfo
  {author} {\bibfnamefont {F.}~\bibnamefont {Giraldo}}, \ and\ \bibinfo
  {author} {\bibfnamefont {D.~S.}\ \bibnamefont {Weiss}},\ }\href {\doibase
  10.1038/s41567-019-0478-8} {\bibfield  {journal} {\bibinfo  {journal} {Nature
  Physics}\ }\textbf {\bibinfo {volume} {15}},\ \bibinfo {pages} {538}
  (\bibinfo {year} {2019})}\BibitemShut {NoStop}%
\bibitem [{\citenamefont {Ebadi}\ \emph {et~al.}(2021)\citenamefont {Ebadi},
  \citenamefont {Wang}, \citenamefont {Levine}, \citenamefont {Keesling},
  \citenamefont {Semeghini}, \citenamefont {Omran}, \citenamefont {Bluvstein},
  \citenamefont {Samajdar}, \citenamefont {Pichler}, \citenamefont {Ho},
  \citenamefont {Choi}, \citenamefont {Sachdev}, \citenamefont {Greiner},
  \citenamefont {Vuleti{\'{c}}},\ and\ \citenamefont {Lukin}}]{Ebadi2021}%
  \BibitemOpen
  \bibfield  {author} {\bibinfo {author} {\bibfnamefont {S.}~\bibnamefont
  {Ebadi}}, \bibinfo {author} {\bibfnamefont {T.~T.}\ \bibnamefont {Wang}},
  \bibinfo {author} {\bibfnamefont {H.}~\bibnamefont {Levine}}, \bibinfo
  {author} {\bibfnamefont {A.}~\bibnamefont {Keesling}}, \bibinfo {author}
  {\bibfnamefont {G.}~\bibnamefont {Semeghini}}, \bibinfo {author}
  {\bibfnamefont {A.}~\bibnamefont {Omran}}, \bibinfo {author} {\bibfnamefont
  {D.}~\bibnamefont {Bluvstein}}, \bibinfo {author} {\bibfnamefont
  {R.}~\bibnamefont {Samajdar}}, \bibinfo {author} {\bibfnamefont
  {H.}~\bibnamefont {Pichler}}, \bibinfo {author} {\bibfnamefont {W.~W.}\
  \bibnamefont {Ho}}, \bibinfo {author} {\bibfnamefont {S.}~\bibnamefont
  {Choi}}, \bibinfo {author} {\bibfnamefont {S.}~\bibnamefont {Sachdev}},
  \bibinfo {author} {\bibfnamefont {M.}~\bibnamefont {Greiner}}, \bibinfo
  {author} {\bibfnamefont {V.}~\bibnamefont {Vuleti{\'{c}}}}, \ and\ \bibinfo
  {author} {\bibfnamefont {M.~D.}\ \bibnamefont {Lukin}},\ }\href {\doibase
  10.1038/s41586-021-03582-4} {\bibfield  {journal} {\bibinfo  {journal}
  {Nature}\ }\textbf {\bibinfo {volume} {595}},\ \bibinfo {pages} {227}
  (\bibinfo {year} {2021})}\BibitemShut {NoStop}%
\bibitem [{\citenamefont {Semeghini}\ \emph {et~al.}(2021)\citenamefont
  {Semeghini}, \citenamefont {Levine}, \citenamefont {Keesling}, \citenamefont
  {Ebadi}, \citenamefont {Wang}, \citenamefont {Bluvstein}, \citenamefont
  {Verresen}, \citenamefont {Pichler}, \citenamefont {Kalinowski},
  \citenamefont {Samajdar}, \citenamefont {Omran}, \citenamefont {Sachdev},
  \citenamefont {Vishwanath}, \citenamefont {Greiner}, \citenamefont
  {Vuletić},\ and\ \citenamefont {Lukin}}]{Semeghini2021}%
  \BibitemOpen
  \bibfield  {author} {\bibinfo {author} {\bibfnamefont {G.}~\bibnamefont
  {Semeghini}}, \bibinfo {author} {\bibfnamefont {H.}~\bibnamefont {Levine}},
  \bibinfo {author} {\bibfnamefont {A.}~\bibnamefont {Keesling}}, \bibinfo
  {author} {\bibfnamefont {S.}~\bibnamefont {Ebadi}}, \bibinfo {author}
  {\bibfnamefont {T.~T.}\ \bibnamefont {Wang}}, \bibinfo {author}
  {\bibfnamefont {D.}~\bibnamefont {Bluvstein}}, \bibinfo {author}
  {\bibfnamefont {R.}~\bibnamefont {Verresen}}, \bibinfo {author}
  {\bibfnamefont {H.}~\bibnamefont {Pichler}}, \bibinfo {author} {\bibfnamefont
  {M.}~\bibnamefont {Kalinowski}}, \bibinfo {author} {\bibfnamefont
  {R.}~\bibnamefont {Samajdar}}, \bibinfo {author} {\bibfnamefont
  {A.}~\bibnamefont {Omran}}, \bibinfo {author} {\bibfnamefont
  {S.}~\bibnamefont {Sachdev}}, \bibinfo {author} {\bibfnamefont
  {A.}~\bibnamefont {Vishwanath}}, \bibinfo {author} {\bibfnamefont
  {M.}~\bibnamefont {Greiner}}, \bibinfo {author} {\bibfnamefont
  {V.}~\bibnamefont {Vuletić}}, \ and\ \bibinfo {author} {\bibfnamefont
  {M.~D.}\ \bibnamefont {Lukin}},\ }\href {\doibase 10.1126/science.abi8794}
  {\bibfield  {journal} {\bibinfo  {journal} {Science}\ }\textbf {\bibinfo
  {volume} {374}},\ \bibinfo {pages} {1242} (\bibinfo {year}
  {2021})}\BibitemShut {NoStop}%
\bibitem [{\citenamefont {Saffman}\ \emph {et~al.}(2010)\citenamefont
  {Saffman}, \citenamefont {Walker},\ and\ \citenamefont
  {M\o{}lmer}}]{Saffman2010}%
  \BibitemOpen
  \bibfield  {author} {\bibinfo {author} {\bibfnamefont {M.}~\bibnamefont
  {Saffman}}, \bibinfo {author} {\bibfnamefont {T.~G.}\ \bibnamefont {Walker}},
  \ and\ \bibinfo {author} {\bibfnamefont {K.}~\bibnamefont {M\o{}lmer}},\
  }\href {\doibase 10.1103/RevModPhys.82.2313} {\bibfield  {journal} {\bibinfo
  {journal} {Rev. Mod. Phys.}\ }\textbf {\bibinfo {volume} {82}},\ \bibinfo
  {pages} {2313} (\bibinfo {year} {2010})}\BibitemShut {NoStop}%
\bibitem [{\citenamefont {Bruzewicz}\ \emph {et~al.}(2019)\citenamefont
  {Bruzewicz}, \citenamefont {Chiaverini}, \citenamefont {McConnell},\ and\
  \citenamefont {Sage}}]{Bruzewicz2019}%
  \BibitemOpen
  \bibfield  {author} {\bibinfo {author} {\bibfnamefont {C.~D.}\ \bibnamefont
  {Bruzewicz}}, \bibinfo {author} {\bibfnamefont {J.}~\bibnamefont
  {Chiaverini}}, \bibinfo {author} {\bibfnamefont {R.}~\bibnamefont
  {McConnell}}, \ and\ \bibinfo {author} {\bibfnamefont {J.~M.}\ \bibnamefont
  {Sage}},\ }\href {\doibase 10.1063/1.5088164} {\bibfield  {journal} {\bibinfo
   {journal} {Applied Physics Reviews}\ }\textbf {\bibinfo {volume} {6}},\
  \bibinfo {pages} {021314} (\bibinfo {year} {2019})}\BibitemShut {NoStop}%
\bibitem [{\citenamefont {Browaeys}\ and\ \citenamefont
  {Lahaye}(2020)}]{Browaeys2020}%
  \BibitemOpen
  \bibfield  {author} {\bibinfo {author} {\bibfnamefont {A.}~\bibnamefont
  {Browaeys}}\ and\ \bibinfo {author} {\bibfnamefont {T.}~\bibnamefont
  {Lahaye}},\ }\href {\doibase 10.1038/s41567-019-0733-z} {\bibfield  {journal}
  {\bibinfo  {journal} {Nature Physics}\ }\textbf {\bibinfo {volume} {16}},\
  \bibinfo {pages} {132} (\bibinfo {year} {2020})}\BibitemShut {NoStop}%
\bibitem [{\citenamefont {Sandvik}(2003)}]{Sandvik2003}%
  \BibitemOpen
  \bibfield  {author} {\bibinfo {author} {\bibfnamefont {A.~W.}\ \bibnamefont
  {Sandvik}},\ }\href {\doibase 10.1103/PhysRevE.68.056701} {\bibfield
  {journal} {\bibinfo  {journal} {Phys. Rev. E}\ }\textbf {\bibinfo {volume}
  {68}},\ \bibinfo {pages} {056701} (\bibinfo {year} {2003})}\BibitemShut
  {NoStop}%
\bibitem [{\citenamefont {Koffel}\ \emph {et~al.}(2012)\citenamefont {Koffel},
  \citenamefont {Lewenstein},\ and\ \citenamefont {Tagliacozzo}}]{Koffel2012}%
  \BibitemOpen
  \bibfield  {author} {\bibinfo {author} {\bibfnamefont {T.}~\bibnamefont
  {Koffel}}, \bibinfo {author} {\bibfnamefont {M.}~\bibnamefont {Lewenstein}},
  \ and\ \bibinfo {author} {\bibfnamefont {L.}~\bibnamefont {Tagliacozzo}},\
  }\href {\doibase 10.1103/PhysRevLett.109.267203} {\bibfield  {journal}
  {\bibinfo  {journal} {Phys. Rev. Lett.}\ }\textbf {\bibinfo {volume} {109}},\
  \bibinfo {pages} {267203} (\bibinfo {year} {2012})}\BibitemShut {NoStop}%
\bibitem [{\citenamefont {Knap}\ \emph {et~al.}(2013)\citenamefont {Knap},
  \citenamefont {Kantian}, \citenamefont {Giamarchi}, \citenamefont {Bloch},
  \citenamefont {Lukin},\ and\ \citenamefont {Demler}}]{Knap2013}%
  \BibitemOpen
  \bibfield  {author} {\bibinfo {author} {\bibfnamefont {M.}~\bibnamefont
  {Knap}}, \bibinfo {author} {\bibfnamefont {A.}~\bibnamefont {Kantian}},
  \bibinfo {author} {\bibfnamefont {T.}~\bibnamefont {Giamarchi}}, \bibinfo
  {author} {\bibfnamefont {I.}~\bibnamefont {Bloch}}, \bibinfo {author}
  {\bibfnamefont {M.~D.}\ \bibnamefont {Lukin}}, \ and\ \bibinfo {author}
  {\bibfnamefont {E.}~\bibnamefont {Demler}},\ }\href {\doibase
  10.1103/PhysRevLett.111.147205} {\bibfield  {journal} {\bibinfo  {journal}
  {Phys. Rev. Lett.}\ }\textbf {\bibinfo {volume} {111}},\ \bibinfo {pages}
  {147205} (\bibinfo {year} {2013})}\BibitemShut {NoStop}%
\bibitem [{\citenamefont {Sun}(2017)}]{Sun2017}%
  \BibitemOpen
  \bibfield  {author} {\bibinfo {author} {\bibfnamefont {G.}~\bibnamefont
  {Sun}},\ }\href {\doibase 10.1103/PhysRevA.96.043621} {\bibfield  {journal}
  {\bibinfo  {journal} {Phys. Rev. A}\ }\textbf {\bibinfo {volume} {96}},\
  \bibinfo {pages} {043621} (\bibinfo {year} {2017})}\BibitemShut {NoStop}%
\bibitem [{\citenamefont {Zhu}\ \emph {et~al.}(2018)\citenamefont {Zhu},
  \citenamefont {Sun}, \citenamefont {You},\ and\ \citenamefont
  {Shi}}]{Zhu2018}%
  \BibitemOpen
  \bibfield  {author} {\bibinfo {author} {\bibfnamefont {Z.}~\bibnamefont
  {Zhu}}, \bibinfo {author} {\bibfnamefont {G.}~\bibnamefont {Sun}}, \bibinfo
  {author} {\bibfnamefont {W.-L.}\ \bibnamefont {You}}, \ and\ \bibinfo
  {author} {\bibfnamefont {D.-N.}\ \bibnamefont {Shi}},\ }\href {\doibase
  10.1103/PhysRevA.98.023607} {\bibfield  {journal} {\bibinfo  {journal} {Phys.
  Rev. A}\ }\textbf {\bibinfo {volume} {98}},\ \bibinfo {pages} {023607}
  (\bibinfo {year} {2018})}\BibitemShut {NoStop}%
\bibitem [{\citenamefont {Sandvik}(2010)}]{Sandvik2010}%
  \BibitemOpen
  \bibfield  {author} {\bibinfo {author} {\bibfnamefont {A.~W.}\ \bibnamefont
  {Sandvik}},\ }\href {\doibase 10.1103/PhysRevLett.104.137204} {\bibfield
  {journal} {\bibinfo  {journal} {Phys. Rev. Lett.}\ }\textbf {\bibinfo
  {volume} {104}},\ \bibinfo {pages} {137204} (\bibinfo {year}
  {2010})}\BibitemShut {NoStop}%
\bibitem [{\citenamefont {Vanderstraeten}\ \emph {et~al.}(2018)\citenamefont
  {Vanderstraeten}, \citenamefont {Van~Damme}, \citenamefont {B\"uchler},\ and\
  \citenamefont {Verstraete}}]{Vanderstraeten2018}%
  \BibitemOpen
  \bibfield  {author} {\bibinfo {author} {\bibfnamefont {L.}~\bibnamefont
  {Vanderstraeten}}, \bibinfo {author} {\bibfnamefont {M.}~\bibnamefont
  {Van~Damme}}, \bibinfo {author} {\bibfnamefont {H.~P.}\ \bibnamefont
  {B\"uchler}}, \ and\ \bibinfo {author} {\bibfnamefont {F.}~\bibnamefont
  {Verstraete}},\ }\href {\doibase 10.1103/PhysRevLett.121.090603} {\bibfield
  {journal} {\bibinfo  {journal} {Phys. Rev. Lett.}\ }\textbf {\bibinfo
  {volume} {121}},\ \bibinfo {pages} {090603} (\bibinfo {year}
  {2018})}\BibitemShut {NoStop}%
\bibitem [{\citenamefont {Fey}\ and\ \citenamefont {Schmidt}(2016)}]{Fey2016}%
  \BibitemOpen
  \bibfield  {author} {\bibinfo {author} {\bibfnamefont {S.}~\bibnamefont
  {Fey}}\ and\ \bibinfo {author} {\bibfnamefont {K.~P.}\ \bibnamefont
  {Schmidt}},\ }\href {\doibase 10.1103/PhysRevB.94.075156} {\bibfield
  {journal} {\bibinfo  {journal} {Phys. Rev. B}\ }\textbf {\bibinfo {volume}
  {94}},\ \bibinfo {pages} {075156} (\bibinfo {year} {2016})}\BibitemShut
  {NoStop}%
\bibitem [{\citenamefont {Adelhardt}\ \emph {et~al.}(2020)\citenamefont
  {Adelhardt}, \citenamefont {Koziol}, \citenamefont {Schellenberger},\ and\
  \citenamefont {Schmidt}}]{Adelhardt2020}%
  \BibitemOpen
  \bibfield  {author} {\bibinfo {author} {\bibfnamefont {P.}~\bibnamefont
  {Adelhardt}}, \bibinfo {author} {\bibfnamefont {J.~A.}\ \bibnamefont
  {Koziol}}, \bibinfo {author} {\bibfnamefont {A.}~\bibnamefont
  {Schellenberger}}, \ and\ \bibinfo {author} {\bibfnamefont {K.~P.}\
  \bibnamefont {Schmidt}},\ }\href {\doibase 10.1103/PhysRevB.102.174424}
  {\bibfield  {journal} {\bibinfo  {journal} {Phys. Rev. B}\ }\textbf {\bibinfo
  {volume} {102}},\ \bibinfo {pages} {174424} (\bibinfo {year}
  {2020})}\BibitemShut {NoStop}%
\bibitem [{\citenamefont {Langheld}\ \emph {et~al.}(2022)\citenamefont
  {Langheld}, \citenamefont {Koziol}, \citenamefont {Adelhardt}, \citenamefont
  {Kapfer},\ and\ \citenamefont {Schmidt}}]{Langheld2022}%
  \BibitemOpen
  \bibfield  {author} {\bibinfo {author} {\bibfnamefont {A.}~\bibnamefont
  {Langheld}}, \bibinfo {author} {\bibfnamefont {J.~A.}\ \bibnamefont
  {Koziol}}, \bibinfo {author} {\bibfnamefont {P.}~\bibnamefont {Adelhardt}},
  \bibinfo {author} {\bibfnamefont {S.~C.}\ \bibnamefont {Kapfer}}, \ and\
  \bibinfo {author} {\bibfnamefont {K.~P.}\ \bibnamefont {Schmidt}},\ }\href
  {\doibase 10.48550/ARXIV.2203.08081} {\enquote {\bibinfo {title} {Scaling at
  quantum phase transitions above the upper critical dimension},}\ } (\bibinfo
  {year} {2022})\BibitemShut {NoStop}%
\bibitem [{\citenamefont {Yang}\ \emph
  {et~al.}(2020{\natexlab{a}})\citenamefont {Yang}, \citenamefont {Yao},\ and\
  \citenamefont {Sandvik}}]{Yang2020}%
  \BibitemOpen
  \bibfield  {author} {\bibinfo {author} {\bibfnamefont {S.}~\bibnamefont
  {Yang}}, \bibinfo {author} {\bibfnamefont {D.-X.}\ \bibnamefont {Yao}}, \
  and\ \bibinfo {author} {\bibfnamefont {A.~W.}\ \bibnamefont {Sandvik}},\
  }\href {\doibase 10.48550/ARXIV.2001.02821} {\enquote {\bibinfo {title}
  {Deconfined quantum criticality in spin-1/2 chains with long-range
  interactions},}\ } (\bibinfo {year} {2020}{\natexlab{a}})\BibitemShut
  {NoStop}%
\bibitem [{\citenamefont {Yusuf}\ \emph {et~al.}(2004)\citenamefont {Yusuf},
  \citenamefont {Joshi},\ and\ \citenamefont {Yang}}]{Yusuf2004}%
  \BibitemOpen
  \bibfield  {author} {\bibinfo {author} {\bibfnamefont {E.}~\bibnamefont
  {Yusuf}}, \bibinfo {author} {\bibfnamefont {A.}~\bibnamefont {Joshi}}, \ and\
  \bibinfo {author} {\bibfnamefont {K.}~\bibnamefont {Yang}},\ }\href {\doibase
  10.1103/PhysRevB.69.144412} {\bibfield  {journal} {\bibinfo  {journal} {Phys.
  Rev. B}\ }\textbf {\bibinfo {volume} {69}},\ \bibinfo {pages} {144412}
  (\bibinfo {year} {2004})}\BibitemShut {NoStop}%
\bibitem [{\citenamefont {Laflorencie}\ \emph {et~al.}(2005)\citenamefont
  {Laflorencie}, \citenamefont {Affleck},\ and\ \citenamefont
  {Berciu}}]{Laflorencie2005}%
  \BibitemOpen
  \bibfield  {author} {\bibinfo {author} {\bibfnamefont {N.}~\bibnamefont
  {Laflorencie}}, \bibinfo {author} {\bibfnamefont {I.}~\bibnamefont
  {Affleck}}, \ and\ \bibinfo {author} {\bibfnamefont {M.}~\bibnamefont
  {Berciu}},\ }\href {\doibase 10.1088/1742-5468/2005/12/p12001} {\bibfield
  {journal} {\bibinfo  {journal} {Journal of Statistical Mechanics: Theory and
  Experiment}\ }\textbf {\bibinfo {volume} {2005}},\ \bibinfo {pages} {P12001}
  (\bibinfo {year} {2005})}\BibitemShut {NoStop}%
\bibitem [{\citenamefont {Zhu}\ and\ \citenamefont {Wang}(2006)}]{Zhu2006}%
  \BibitemOpen
  \bibfield  {author} {\bibinfo {author} {\bibfnamefont {R.-G.}\ \bibnamefont
  {Zhu}}\ and\ \bibinfo {author} {\bibfnamefont {A.-M.}\ \bibnamefont {Wang}},\
  }\href {\doibase 10.1103/PhysRevB.74.012406} {\bibfield  {journal} {\bibinfo
  {journal} {Phys. Rev. B}\ }\textbf {\bibinfo {volume} {74}},\ \bibinfo
  {pages} {012406} (\bibinfo {year} {2006})}\BibitemShut {NoStop}%
\bibitem [{\citenamefont {Li}\ and\ \citenamefont {Wang}(2015)}]{Li2015}%
  \BibitemOpen
  \bibfield  {author} {\bibinfo {author} {\bibfnamefont {Z.-H.}\ \bibnamefont
  {Li}}\ and\ \bibinfo {author} {\bibfnamefont {A.-M.}\ \bibnamefont {Wang}},\
  }\href {\doibase 10.1103/PhysRevB.91.235110} {\bibfield  {journal} {\bibinfo
  {journal} {Phys. Rev. B}\ }\textbf {\bibinfo {volume} {91}},\ \bibinfo
  {pages} {235110} (\bibinfo {year} {2015})}\BibitemShut {NoStop}%
\bibitem [{\citenamefont {Tang}\ and\ \citenamefont
  {Sandvik}(2015)}]{Tang2015}%
  \BibitemOpen
  \bibfield  {author} {\bibinfo {author} {\bibfnamefont {Y.}~\bibnamefont
  {Tang}}\ and\ \bibinfo {author} {\bibfnamefont {A.~W.}\ \bibnamefont
  {Sandvik}},\ }\href {\doibase 10.1103/PhysRevB.92.184425} {\bibfield
  {journal} {\bibinfo  {journal} {Phys. Rev. B}\ }\textbf {\bibinfo {volume}
  {92}},\ \bibinfo {pages} {184425} (\bibinfo {year} {2015})}\BibitemShut
  {NoStop}%
\bibitem [{\citenamefont {Gong}\ \emph {et~al.}(2016)\citenamefont {Gong},
  \citenamefont {Maghrebi}, \citenamefont {Hu}, \citenamefont {Foss-Feig},
  \citenamefont {Richerme}, \citenamefont {Monroe},\ and\ \citenamefont
  {Gorshkov}}]{Gong2016}%
  \BibitemOpen
  \bibfield  {author} {\bibinfo {author} {\bibfnamefont {Z.-X.}\ \bibnamefont
  {Gong}}, \bibinfo {author} {\bibfnamefont {M.~F.}\ \bibnamefont {Maghrebi}},
  \bibinfo {author} {\bibfnamefont {A.}~\bibnamefont {Hu}}, \bibinfo {author}
  {\bibfnamefont {M.}~\bibnamefont {Foss-Feig}}, \bibinfo {author}
  {\bibfnamefont {P.}~\bibnamefont {Richerme}}, \bibinfo {author}
  {\bibfnamefont {C.}~\bibnamefont {Monroe}}, \ and\ \bibinfo {author}
  {\bibfnamefont {A.~V.}\ \bibnamefont {Gorshkov}},\ }\href {\doibase
  10.1103/PhysRevB.93.205115} {\bibfield  {journal} {\bibinfo  {journal} {Phys.
  Rev. B}\ }\textbf {\bibinfo {volume} {93}},\ \bibinfo {pages} {205115}
  (\bibinfo {year} {2016})}\BibitemShut {NoStop}%
\bibitem [{\citenamefont {Maghrebi}\ \emph {et~al.}(2017)\citenamefont
  {Maghrebi}, \citenamefont {Gong},\ and\ \citenamefont
  {Gorshkov}}]{Maghrebi2017}%
  \BibitemOpen
  \bibfield  {author} {\bibinfo {author} {\bibfnamefont {M.~F.}\ \bibnamefont
  {Maghrebi}}, \bibinfo {author} {\bibfnamefont {Z.-X.}\ \bibnamefont {Gong}},
  \ and\ \bibinfo {author} {\bibfnamefont {A.~V.}\ \bibnamefont {Gorshkov}},\
  }\href {\doibase 10.1103/PhysRevLett.119.023001} {\bibfield  {journal}
  {\bibinfo  {journal} {Phys. Rev. Lett.}\ }\textbf {\bibinfo {volume} {119}},\
  \bibinfo {pages} {023001} (\bibinfo {year} {2017})}\BibitemShut {NoStop}%
\bibitem [{\citenamefont {Ren}\ \emph {et~al.}(2020)\citenamefont {Ren},
  \citenamefont {You},\ and\ \citenamefont {Ole\ifmmode~\acute{s}\else
  \'{s}\fi{}}}]{Ren2020}%
  \BibitemOpen
  \bibfield  {author} {\bibinfo {author} {\bibfnamefont {J.}~\bibnamefont
  {Ren}}, \bibinfo {author} {\bibfnamefont {W.-L.}\ \bibnamefont {You}}, \ and\
  \bibinfo {author} {\bibfnamefont {A.~M.}\ \bibnamefont
  {Ole\ifmmode~\acute{s}\else \'{s}\fi{}}},\ }\href {\doibase
  10.1103/PhysRevB.102.024425} {\bibfield  {journal} {\bibinfo  {journal}
  {Phys. Rev. B}\ }\textbf {\bibinfo {volume} {102}},\ \bibinfo {pages}
  {024425} (\bibinfo {year} {2020})}\BibitemShut {NoStop}%
\bibitem [{\citenamefont {Vodola}\ \emph {et~al.}(2014)\citenamefont {Vodola},
  \citenamefont {Lepori}, \citenamefont {Ercolessi}, \citenamefont {Gorshkov},\
  and\ \citenamefont {Pupillo}}]{Vodola2014}%
  \BibitemOpen
  \bibfield  {author} {\bibinfo {author} {\bibfnamefont {D.}~\bibnamefont
  {Vodola}}, \bibinfo {author} {\bibfnamefont {L.}~\bibnamefont {Lepori}},
  \bibinfo {author} {\bibfnamefont {E.}~\bibnamefont {Ercolessi}}, \bibinfo
  {author} {\bibfnamefont {A.~V.}\ \bibnamefont {Gorshkov}}, \ and\ \bibinfo
  {author} {\bibfnamefont {G.}~\bibnamefont {Pupillo}},\ }\href {\doibase
  10.1103/PhysRevLett.113.156402} {\bibfield  {journal} {\bibinfo  {journal}
  {Phys. Rev. Lett.}\ }\textbf {\bibinfo {volume} {113}},\ \bibinfo {pages}
  {156402} (\bibinfo {year} {2014})}\BibitemShut {NoStop}%
\bibitem [{\citenamefont {Vodola}\ \emph {et~al.}(2015)\citenamefont {Vodola},
  \citenamefont {Lepori}, \citenamefont {Ercolessi},\ and\ \citenamefont
  {Pupillo}}]{Vodola2015}%
  \BibitemOpen
  \bibfield  {author} {\bibinfo {author} {\bibfnamefont {D.}~\bibnamefont
  {Vodola}}, \bibinfo {author} {\bibfnamefont {L.}~\bibnamefont {Lepori}},
  \bibinfo {author} {\bibfnamefont {E.}~\bibnamefont {Ercolessi}}, \ and\
  \bibinfo {author} {\bibfnamefont {G.}~\bibnamefont {Pupillo}},\ }\href
  {\doibase 10.1088/1367-2630/18/1/015001} {\bibfield  {journal} {\bibinfo
  {journal} {New Journal of Physics}\ }\textbf {\bibinfo {volume} {18}},\
  \bibinfo {pages} {015001} (\bibinfo {year} {2015})}\BibitemShut {NoStop}%
\bibitem [{\citenamefont {Maity}\ \emph {et~al.}(2019)\citenamefont {Maity},
  \citenamefont {Bhattacharya},\ and\ \citenamefont {Dutta}}]{Maity2019}%
  \BibitemOpen
  \bibfield  {author} {\bibinfo {author} {\bibfnamefont {S.}~\bibnamefont
  {Maity}}, \bibinfo {author} {\bibfnamefont {U.}~\bibnamefont {Bhattacharya}},
  \ and\ \bibinfo {author} {\bibfnamefont {A.}~\bibnamefont {Dutta}},\ }\href
  {\doibase 10.1088/1751-8121/ab5634} {\bibfield  {journal} {\bibinfo
  {journal} {Journal of Physics A: Mathematical and Theoretical}\ }\textbf
  {\bibinfo {volume} {53}},\ \bibinfo {pages} {013001} (\bibinfo {year}
  {2019})}\BibitemShut {NoStop}%
\bibitem [{\citenamefont {Sadhukhan}\ and\ \citenamefont
  {Dziarmaga}(2021)}]{Sadhukhan2021}%
  \BibitemOpen
  \bibfield  {author} {\bibinfo {author} {\bibfnamefont {D.}~\bibnamefont
  {Sadhukhan}}\ and\ \bibinfo {author} {\bibfnamefont {J.}~\bibnamefont
  {Dziarmaga}},\ }\href {\doibase 10.48550/ARXIV.2107.02508} {\enquote
  {\bibinfo {title} {Is there a correlation length in a model with long-range
  interactions?}}\ } (\bibinfo {year} {2021})\BibitemShut {NoStop}%
\bibitem [{\citenamefont {Samajdar}\ \emph {et~al.}(2021)\citenamefont
  {Samajdar}, \citenamefont {Ho}, \citenamefont {Pichler}, \citenamefont
  {Lukin},\ and\ \citenamefont {Sachdev}}]{Samajdar2021}%
  \BibitemOpen
  \bibfield  {author} {\bibinfo {author} {\bibfnamefont {R.}~\bibnamefont
  {Samajdar}}, \bibinfo {author} {\bibfnamefont {W.~W.}\ \bibnamefont {Ho}},
  \bibinfo {author} {\bibfnamefont {H.}~\bibnamefont {Pichler}}, \bibinfo
  {author} {\bibfnamefont {M.~D.}\ \bibnamefont {Lukin}}, \ and\ \bibinfo
  {author} {\bibfnamefont {S.}~\bibnamefont {Sachdev}},\ }\href {\doibase
  10.1073/pnas.2015785118} {\bibfield  {journal} {\bibinfo  {journal}
  {Proceedings of the National Academy of Sciences}\ }\textbf {\bibinfo
  {volume} {118}},\ \bibinfo {pages} {e2015785118} (\bibinfo {year}
  {2021})}\BibitemShut {NoStop}%
\bibitem [{\citenamefont {Verresen}\ \emph {et~al.}(2021)\citenamefont
  {Verresen}, \citenamefont {Lukin},\ and\ \citenamefont
  {Vishwanath}}]{Verresen2021}%
  \BibitemOpen
  \bibfield  {author} {\bibinfo {author} {\bibfnamefont {R.}~\bibnamefont
  {Verresen}}, \bibinfo {author} {\bibfnamefont {M.~D.}\ \bibnamefont {Lukin}},
  \ and\ \bibinfo {author} {\bibfnamefont {A.}~\bibnamefont {Vishwanath}},\
  }\href {\doibase 10.1103/PhysRevX.11.031005} {\bibfield  {journal} {\bibinfo
  {journal} {Phys. Rev. X}\ }\textbf {\bibinfo {volume} {11}},\ \bibinfo
  {pages} {031005} (\bibinfo {year} {2021})}\BibitemShut {NoStop}%
\bibitem [{\citenamefont {Liu}\ \emph {et~al.}(2022)\citenamefont {Liu},
  \citenamefont {Yang}, \citenamefont {Bienias}, \citenamefont {Iadecola},\
  and\ \citenamefont {Gorshkov}}]{Liu2022}%
  \BibitemOpen
  \bibfield  {author} {\bibinfo {author} {\bibfnamefont {F.}~\bibnamefont
  {Liu}}, \bibinfo {author} {\bibfnamefont {Z.-C.}\ \bibnamefont {Yang}},
  \bibinfo {author} {\bibfnamefont {P.}~\bibnamefont {Bienias}}, \bibinfo
  {author} {\bibfnamefont {T.}~\bibnamefont {Iadecola}}, \ and\ \bibinfo
  {author} {\bibfnamefont {A.~V.}\ \bibnamefont {Gorshkov}},\ }\href {\doibase
  10.1103/PhysRevLett.128.013603} {\bibfield  {journal} {\bibinfo  {journal}
  {Phys. Rev. Lett.}\ }\textbf {\bibinfo {volume} {128}},\ \bibinfo {pages}
  {013603} (\bibinfo {year} {2022})}\BibitemShut {NoStop}%
\bibitem [{\citenamefont {Humeniuk}(2016)}]{Humeniuk2016}%
  \BibitemOpen
  \bibfield  {author} {\bibinfo {author} {\bibfnamefont {S.}~\bibnamefont
  {Humeniuk}},\ }\href {\doibase 10.1103/PhysRevB.93.104412} {\bibfield
  {journal} {\bibinfo  {journal} {Phys. Rev. B}\ }\textbf {\bibinfo {volume}
  {93}},\ \bibinfo {pages} {104412} (\bibinfo {year} {2016})}\BibitemShut
  {NoStop}%
\bibitem [{\citenamefont {Fey}\ \emph {et~al.}(2019)\citenamefont {Fey},
  \citenamefont {Kapfer},\ and\ \citenamefont {Schmidt}}]{Fey2019}%
  \BibitemOpen
  \bibfield  {author} {\bibinfo {author} {\bibfnamefont {S.}~\bibnamefont
  {Fey}}, \bibinfo {author} {\bibfnamefont {S.~C.}\ \bibnamefont {Kapfer}}, \
  and\ \bibinfo {author} {\bibfnamefont {K.~P.}\ \bibnamefont {Schmidt}},\
  }\href {\doibase 10.1103/PhysRevLett.122.017203} {\bibfield  {journal}
  {\bibinfo  {journal} {Phys. Rev. Lett.}\ }\textbf {\bibinfo {volume} {122}},\
  \bibinfo {pages} {017203} (\bibinfo {year} {2019})}\BibitemShut {NoStop}%
\bibitem [{\citenamefont {Koziol}\ \emph {et~al.}(2021)\citenamefont {Koziol},
  \citenamefont {Langheld}, \citenamefont {Kapfer},\ and\ \citenamefont
  {Schmidt}}]{Koziol2021}%
  \BibitemOpen
  \bibfield  {author} {\bibinfo {author} {\bibfnamefont {J.~A.}\ \bibnamefont
  {Koziol}}, \bibinfo {author} {\bibfnamefont {A.}~\bibnamefont {Langheld}},
  \bibinfo {author} {\bibfnamefont {S.~C.}\ \bibnamefont {Kapfer}}, \ and\
  \bibinfo {author} {\bibfnamefont {K.~P.}\ \bibnamefont {Schmidt}},\ }\href
  {\doibase 10.1103/PhysRevB.103.245135} {\bibfield  {journal} {\bibinfo
  {journal} {Phys. Rev. B}\ }\textbf {\bibinfo {volume} {103}},\ \bibinfo
  {pages} {245135} (\bibinfo {year} {2021})}\BibitemShut {NoStop}%
\bibitem [{\citenamefont {Dutta}\ and\ \citenamefont
  {Bhattacharjee}(2001)}]{Dutta2001}%
  \BibitemOpen
  \bibfield  {author} {\bibinfo {author} {\bibfnamefont {A.}~\bibnamefont
  {Dutta}}\ and\ \bibinfo {author} {\bibfnamefont {J.~K.}\ \bibnamefont
  {Bhattacharjee}},\ }\href {\doibase 10.1103/PhysRevB.64.184106} {\bibfield
  {journal} {\bibinfo  {journal} {Phys. Rev. B}\ }\textbf {\bibinfo {volume}
  {64}},\ \bibinfo {pages} {184106} (\bibinfo {year} {2001})}\BibitemShut
  {NoStop}%
\bibitem [{\citenamefont {Sak}(1973)}]{Sak1973}%
  \BibitemOpen
  \bibfield  {author} {\bibinfo {author} {\bibfnamefont {J.}~\bibnamefont
  {Sak}},\ }\href {\doibase 10.1103/PhysRevB.8.281} {\bibfield  {journal}
  {\bibinfo  {journal} {Phys. Rev. B}\ }\textbf {\bibinfo {volume} {8}},\
  \bibinfo {pages} {281} (\bibinfo {year} {1973})}\BibitemShut {NoStop}%
\bibitem [{\citenamefont {Defenu}\ \emph {et~al.}(2017)\citenamefont {Defenu},
  \citenamefont {Trombettoni},\ and\ \citenamefont {Ruffo}}]{Defenu2017}%
  \BibitemOpen
  \bibfield  {author} {\bibinfo {author} {\bibfnamefont {N.}~\bibnamefont
  {Defenu}}, \bibinfo {author} {\bibfnamefont {A.}~\bibnamefont {Trombettoni}},
  \ and\ \bibinfo {author} {\bibfnamefont {S.}~\bibnamefont {Ruffo}},\ }\href
  {\doibase 10.1103/PhysRevB.96.104432} {\bibfield  {journal} {\bibinfo
  {journal} {Phys. Rev. B}\ }\textbf {\bibinfo {volume} {96}},\ \bibinfo
  {pages} {104432} (\bibinfo {year} {2017})}\BibitemShut {NoStop}%
\bibitem [{\citenamefont {Behan}\ \emph
  {et~al.}(2017{\natexlab{a}})\citenamefont {Behan}, \citenamefont {Rastelli},
  \citenamefont {Rychkov},\ and\ \citenamefont {Zan}}]{Behan2017a}%
  \BibitemOpen
  \bibfield  {author} {\bibinfo {author} {\bibfnamefont {C.}~\bibnamefont
  {Behan}}, \bibinfo {author} {\bibfnamefont {L.}~\bibnamefont {Rastelli}},
  \bibinfo {author} {\bibfnamefont {S.}~\bibnamefont {Rychkov}}, \ and\
  \bibinfo {author} {\bibfnamefont {B.}~\bibnamefont {Zan}},\ }\href {\doibase
  10.1103/PhysRevLett.118.241601} {\bibfield  {journal} {\bibinfo  {journal}
  {Phys. Rev. Lett.}\ }\textbf {\bibinfo {volume} {118}},\ \bibinfo {pages}
  {241601} (\bibinfo {year} {2017}{\natexlab{a}})}\BibitemShut {NoStop}%
\bibitem [{\citenamefont {Behan}\ \emph
  {et~al.}(2017{\natexlab{b}})\citenamefont {Behan}, \citenamefont {Rastelli},
  \citenamefont {Rychkov},\ and\ \citenamefont {Zan}}]{Behan2017b}%
  \BibitemOpen
  \bibfield  {author} {\bibinfo {author} {\bibfnamefont {C.}~\bibnamefont
  {Behan}}, \bibinfo {author} {\bibfnamefont {L.}~\bibnamefont {Rastelli}},
  \bibinfo {author} {\bibfnamefont {S.}~\bibnamefont {Rychkov}}, \ and\
  \bibinfo {author} {\bibfnamefont {B.}~\bibnamefont {Zan}},\ }\href {\doibase
  10.1088/1751-8121/aa8099} {\bibfield  {journal} {\bibinfo  {journal} {J.
  Phys. A}\ }\textbf {\bibinfo {volume} {50}},\ \bibinfo {pages} {354002}
  (\bibinfo {year} {2017}{\natexlab{b}})}\BibitemShut {NoStop}%
\bibitem [{\citenamefont {Defenu}\ \emph {et~al.}(2020)\citenamefont {Defenu},
  \citenamefont {Codello}, \citenamefont {Ruffo},\ and\ \citenamefont
  {Trombettoni}}]{Defenu2020}%
  \BibitemOpen
  \bibfield  {author} {\bibinfo {author} {\bibfnamefont {N.}~\bibnamefont
  {Defenu}}, \bibinfo {author} {\bibfnamefont {A.}~\bibnamefont {Codello}},
  \bibinfo {author} {\bibfnamefont {S.}~\bibnamefont {Ruffo}}, \ and\ \bibinfo
  {author} {\bibfnamefont {A.}~\bibnamefont {Trombettoni}},\ }\href {\doibase
  10.1088/1751-8121/ab6a6c} {\bibfield  {journal} {\bibinfo  {journal} {J.
  Phys. A}\ }\textbf {\bibinfo {volume} {53}},\ \bibinfo {pages} {143001}
  (\bibinfo {year} {2020})}\BibitemShut {NoStop}%
\bibitem [{\citenamefont {Mermin}\ and\ \citenamefont
  {Wagner}(1966)}]{Mermin1966}%
  \BibitemOpen
  \bibfield  {author} {\bibinfo {author} {\bibfnamefont {N.~D.}\ \bibnamefont
  {Mermin}}\ and\ \bibinfo {author} {\bibfnamefont {H.}~\bibnamefont
  {Wagner}},\ }\href {\doibase 10.1103/PhysRevLett.17.1133} {\bibfield
  {journal} {\bibinfo  {journal} {Phys. Rev. Lett.}\ }\textbf {\bibinfo
  {volume} {17}},\ \bibinfo {pages} {1133} (\bibinfo {year}
  {1966})}\BibitemShut {NoStop}%
\bibitem [{\citenamefont {Hohenberg}(1967)}]{Hohenberg1967}%
  \BibitemOpen
  \bibfield  {author} {\bibinfo {author} {\bibfnamefont {P.~C.}\ \bibnamefont
  {Hohenberg}},\ }\href {\doibase 10.1103/PhysRev.158.383} {\bibfield
  {journal} {\bibinfo  {journal} {Phys. Rev.}\ }\textbf {\bibinfo {volume}
  {158}},\ \bibinfo {pages} {383} (\bibinfo {year} {1967})}\BibitemShut
  {NoStop}%
\bibitem [{\citenamefont {Coleman}(1973)}]{Coleman1973}%
  \BibitemOpen
  \bibfield  {author} {\bibinfo {author} {\bibfnamefont {S.}~\bibnamefont
  {Coleman}},\ }\href {\doibase 10.1007/BF01646487} {\bibfield  {journal}
  {\bibinfo  {journal} {Communications in Mathematical Physics}\ }\textbf
  {\bibinfo {volume} {31}},\ \bibinfo {pages} {259} (\bibinfo {year}
  {1973})}\BibitemShut {NoStop}%
\bibitem [{\citenamefont {Bruno}(2001)}]{Bruno2001}%
  \BibitemOpen
  \bibfield  {author} {\bibinfo {author} {\bibfnamefont {P.}~\bibnamefont
  {Bruno}},\ }\href {\doibase 10.1103/PhysRevLett.87.137203} {\bibfield
  {journal} {\bibinfo  {journal} {Phys. Rev. Lett.}\ }\textbf {\bibinfo
  {volume} {87}},\ \bibinfo {pages} {137203} (\bibinfo {year}
  {2001})}\BibitemShut {NoStop}%
\bibitem [{\citenamefont {Yang}\ and\ \citenamefont
  {Feiguin}(2021)}]{Yang2021}%
  \BibitemOpen
  \bibfield  {author} {\bibinfo {author} {\bibfnamefont {L.}~\bibnamefont
  {Yang}}\ and\ \bibinfo {author} {\bibfnamefont {A.~E.}\ \bibnamefont
  {Feiguin}},\ }\href {\doibase 10.21468/SciPostPhys.10.5.110} {\bibfield
  {journal} {\bibinfo  {journal} {SciPost Phys.}\ }\textbf {\bibinfo {volume}
  {10}},\ \bibinfo {pages} {110} (\bibinfo {year} {2021})}\BibitemShut
  {NoStop}%
\bibitem [{\citenamefont {Yang}\ \emph
  {et~al.}(2020{\natexlab{b}})\citenamefont {Yang}, \citenamefont {Weinberg},\
  and\ \citenamefont {Feiguin}}]{Yang2022}%
  \BibitemOpen
  \bibfield  {author} {\bibinfo {author} {\bibfnamefont {L.}~\bibnamefont
  {Yang}}, \bibinfo {author} {\bibfnamefont {P.}~\bibnamefont {Weinberg}}, \
  and\ \bibinfo {author} {\bibfnamefont {A.~E.}\ \bibnamefont {Feiguin}},\
  }\href {\doibase 10.48550/ARXIV.2012.14908} {\enquote {\bibinfo {title}
  {Topological to magnetically ordered quantum phase transition in
  antiferromagnetic spin ladders with long-range interactions},}\ } (\bibinfo
  {year} {2020}{\natexlab{b}})\BibitemShut {NoStop}%
\bibitem [{\citenamefont {Vishwanath}\ \emph {et~al.}(2004)\citenamefont
  {Vishwanath}, \citenamefont {Balents},\ and\ \citenamefont
  {Senthil}}]{Vishwanath2004}%
  \BibitemOpen
  \bibfield  {author} {\bibinfo {author} {\bibfnamefont {A.}~\bibnamefont
  {Vishwanath}}, \bibinfo {author} {\bibfnamefont {L.}~\bibnamefont {Balents}},
  \ and\ \bibinfo {author} {\bibfnamefont {T.}~\bibnamefont {Senthil}},\ }\href
  {\doibase 10.1103/PhysRevB.69.224416} {\bibfield  {journal} {\bibinfo
  {journal} {Phys. Rev. B}\ }\textbf {\bibinfo {volume} {69}},\ \bibinfo
  {pages} {224416} (\bibinfo {year} {2004})}\BibitemShut {NoStop}%
\bibitem [{\citenamefont {Senthil}\ \emph
  {et~al.}(2004{\natexlab{a}})\citenamefont {Senthil}, \citenamefont
  {Vishwanath}, \citenamefont {Balents}, \citenamefont {Sachdev},\ and\
  \citenamefont {Fisher}}]{Senthil2004a}%
  \BibitemOpen
  \bibfield  {author} {\bibinfo {author} {\bibfnamefont {T.}~\bibnamefont
  {Senthil}}, \bibinfo {author} {\bibfnamefont {A.}~\bibnamefont {Vishwanath}},
  \bibinfo {author} {\bibfnamefont {L.}~\bibnamefont {Balents}}, \bibinfo
  {author} {\bibfnamefont {S.}~\bibnamefont {Sachdev}}, \ and\ \bibinfo
  {author} {\bibfnamefont {M.~P.~A.}\ \bibnamefont {Fisher}},\ }\href {\doibase
  10.1126/science.1091806} {\bibfield  {journal} {\bibinfo  {journal}
  {Science}\ }\textbf {\bibinfo {volume} {303}},\ \bibinfo {pages} {1490}
  (\bibinfo {year} {2004}{\natexlab{a}})}\BibitemShut {NoStop}%
\bibitem [{\citenamefont {Senthil}\ \emph
  {et~al.}(2004{\natexlab{b}})\citenamefont {Senthil}, \citenamefont {Balents},
  \citenamefont {Sachdev}, \citenamefont {Vishwanath},\ and\ \citenamefont
  {Fisher}}]{Senthil2004b}%
  \BibitemOpen
  \bibfield  {author} {\bibinfo {author} {\bibfnamefont {T.}~\bibnamefont
  {Senthil}}, \bibinfo {author} {\bibfnamefont {L.}~\bibnamefont {Balents}},
  \bibinfo {author} {\bibfnamefont {S.}~\bibnamefont {Sachdev}}, \bibinfo
  {author} {\bibfnamefont {A.}~\bibnamefont {Vishwanath}}, \ and\ \bibinfo
  {author} {\bibfnamefont {M.~P.~A.}\ \bibnamefont {Fisher}},\ }\href {\doibase
  10.1103/PhysRevB.70.144407} {\bibfield  {journal} {\bibinfo  {journal} {Phys.
  Rev. B}\ }\textbf {\bibinfo {volume} {70}},\ \bibinfo {pages} {144407}
  (\bibinfo {year} {2004}{\natexlab{b}})}\BibitemShut {NoStop}%
\bibitem [{\citenamefont {Senthil}\ \emph {et~al.}(2005)\citenamefont
  {Senthil}, \citenamefont {Balents}, \citenamefont {Sachdev}, \citenamefont
  {Vishwanath},\ and\ \citenamefont {P.~A.~Fisher}}]{Senthil2005}%
  \BibitemOpen
  \bibfield  {author} {\bibinfo {author} {\bibfnamefont {T.}~\bibnamefont
  {Senthil}}, \bibinfo {author} {\bibfnamefont {L.}~\bibnamefont {Balents}},
  \bibinfo {author} {\bibfnamefont {S.}~\bibnamefont {Sachdev}}, \bibinfo
  {author} {\bibfnamefont {A.}~\bibnamefont {Vishwanath}}, \ and\ \bibinfo
  {author} {\bibfnamefont {M.}~\bibnamefont {P.~A.~Fisher}},\ }\href {\doibase
  10.1143/JPSJS.74S.1} {\bibfield  {journal} {\bibinfo  {journal} {Journal of
  the Physical Society of Japan}\ }\textbf {\bibinfo {volume} {74}},\ \bibinfo
  {pages} {1} (\bibinfo {year} {2005})}\BibitemShut {NoStop}%
\bibitem [{\citenamefont {Kim}\ \emph {et~al.}(2000)\citenamefont {Kim},
  \citenamefont {F\'ath}, \citenamefont {S\'olyom},\ and\ \citenamefont
  {Scalapino}}]{Kim2000}%
  \BibitemOpen
  \bibfield  {author} {\bibinfo {author} {\bibfnamefont {E.~H.}\ \bibnamefont
  {Kim}}, \bibinfo {author} {\bibfnamefont {G.}~\bibnamefont {F\'ath}},
  \bibinfo {author} {\bibfnamefont {J.}~\bibnamefont {S\'olyom}}, \ and\
  \bibinfo {author} {\bibfnamefont {D.~J.}\ \bibnamefont {Scalapino}},\ }\href
  {\doibase 10.1103/PhysRevB.62.14965} {\bibfield  {journal} {\bibinfo
  {journal} {Phys. Rev. B}\ }\textbf {\bibinfo {volume} {62}},\ \bibinfo
  {pages} {14965} (\bibinfo {year} {2000})}\BibitemShut {NoStop}%
\bibitem [{\citenamefont {Pollmann}\ \emph {et~al.}(2012)\citenamefont
  {Pollmann}, \citenamefont {Berg}, \citenamefont {Turner},\ and\ \citenamefont
  {Oshikawa}}]{Pollmann2012}%
  \BibitemOpen
  \bibfield  {author} {\bibinfo {author} {\bibfnamefont {F.}~\bibnamefont
  {Pollmann}}, \bibinfo {author} {\bibfnamefont {E.}~\bibnamefont {Berg}},
  \bibinfo {author} {\bibfnamefont {A.~M.}\ \bibnamefont {Turner}}, \ and\
  \bibinfo {author} {\bibfnamefont {M.}~\bibnamefont {Oshikawa}},\ }\href
  {\doibase 10.1103/PhysRevB.85.075125} {\bibfield  {journal} {\bibinfo
  {journal} {Phys. Rev. B}\ }\textbf {\bibinfo {volume} {85}},\ \bibinfo
  {pages} {075125} (\bibinfo {year} {2012})}\BibitemShut {NoStop}%
\bibitem [{\citenamefont {Takada}\ and\ \citenamefont
  {Watanabe}(1992)}]{Takada1992}%
  \BibitemOpen
  \bibfield  {author} {\bibinfo {author} {\bibfnamefont {S.}~\bibnamefont
  {Takada}}\ and\ \bibinfo {author} {\bibfnamefont {H.}~\bibnamefont
  {Watanabe}},\ }\href {\doibase 10.1143/JPSJ.61.39} {\bibfield  {journal}
  {\bibinfo  {journal} {Journal of the Physical Society of Japan}\ }\textbf
  {\bibinfo {volume} {61}},\ \bibinfo {pages} {39} (\bibinfo {year}
  {1992})}\BibitemShut {NoStop}%
\bibitem [{\citenamefont {Watanabe}(1995)}]{Watanabe1995}%
  \BibitemOpen
  \bibfield  {author} {\bibinfo {author} {\bibfnamefont {H.}~\bibnamefont
  {Watanabe}},\ }\href {\doibase 10.1103/PhysRevB.52.12508} {\bibfield
  {journal} {\bibinfo  {journal} {Phys. Rev. B}\ }\textbf {\bibinfo {volume}
  {52}},\ \bibinfo {pages} {12508} (\bibinfo {year} {1995})}\BibitemShut
  {NoStop}%
\bibitem [{\citenamefont {Nishiyama}\ \emph {et~al.}(1995)\citenamefont
  {Nishiyama}, \citenamefont {Hatano},\ and\ \citenamefont
  {Suzuki}}]{Nishiyama1995}%
  \BibitemOpen
  \bibfield  {author} {\bibinfo {author} {\bibfnamefont {Y.}~\bibnamefont
  {Nishiyama}}, \bibinfo {author} {\bibfnamefont {N.}~\bibnamefont {Hatano}}, \
  and\ \bibinfo {author} {\bibfnamefont {M.}~\bibnamefont {Suzuki}},\ }\href
  {\doibase 10.1143/JPSJ.64.1967} {\bibfield  {journal} {\bibinfo  {journal}
  {Journal of the Physical Society of Japan}\ }\textbf {\bibinfo {volume}
  {64}},\ \bibinfo {pages} {1967} (\bibinfo {year} {1995})}\BibitemShut
  {NoStop}%
\bibitem [{\citenamefont {White}(1996)}]{White1996}%
  \BibitemOpen
  \bibfield  {author} {\bibinfo {author} {\bibfnamefont {S.~R.}\ \bibnamefont
  {White}},\ }\href {\doibase 10.1103/PhysRevB.53.52} {\bibfield  {journal}
  {\bibinfo  {journal} {Phys. Rev. B}\ }\textbf {\bibinfo {volume} {53}},\
  \bibinfo {pages} {52} (\bibinfo {year} {1996})}\BibitemShut {NoStop}%
\bibitem [{\citenamefont {Nambu}(1960)}]{Nambu1960}%
  \BibitemOpen
  \bibfield  {author} {\bibinfo {author} {\bibfnamefont {Y.}~\bibnamefont
  {Nambu}},\ }\href {\doibase 10.1103/PhysRev.117.648} {\bibfield  {journal}
  {\bibinfo  {journal} {Phys. Rev.}\ }\textbf {\bibinfo {volume} {117}},\
  \bibinfo {pages} {648} (\bibinfo {year} {1960})}\BibitemShut {NoStop}%
\bibitem [{\citenamefont {Goldstone}(1961)}]{Goldstone1961}%
  \BibitemOpen
  \bibfield  {author} {\bibinfo {author} {\bibfnamefont {J.}~\bibnamefont
  {Goldstone}},\ }\href {\doibase 10.1007/BF02812722} {\bibfield  {journal}
  {\bibinfo  {journal} {Il Nuovo Cimento (1955-1965)}\ }\textbf {\bibinfo
  {volume} {19}},\ \bibinfo {pages} {154} (\bibinfo {year} {1961})}\BibitemShut
  {NoStop}%
\bibitem [{\citenamefont {Goldstone}\ \emph {et~al.}(1962)\citenamefont
  {Goldstone}, \citenamefont {Salam},\ and\ \citenamefont
  {Weinberg}}]{Goldstone1962}%
  \BibitemOpen
  \bibfield  {author} {\bibinfo {author} {\bibfnamefont {J.}~\bibnamefont
  {Goldstone}}, \bibinfo {author} {\bibfnamefont {A.}~\bibnamefont {Salam}}, \
  and\ \bibinfo {author} {\bibfnamefont {S.}~\bibnamefont {Weinberg}},\ }\href
  {\doibase 10.1103/PhysRev.127.965} {\bibfield  {journal} {\bibinfo  {journal}
  {Phys. Rev.}\ }\textbf {\bibinfo {volume} {127}},\ \bibinfo {pages} {965}
  (\bibinfo {year} {1962})}\BibitemShut {NoStop}%
\bibitem [{\citenamefont {Sachdev}\ and\ \citenamefont
  {Bhatt}(1990)}]{Sachdev1990}%
  \BibitemOpen
  \bibfield  {author} {\bibinfo {author} {\bibfnamefont {S.}~\bibnamefont
  {Sachdev}}\ and\ \bibinfo {author} {\bibfnamefont {R.~N.}\ \bibnamefont
  {Bhatt}},\ }\href {\doibase 10.1103/PhysRevB.41.9323} {\bibfield  {journal}
  {\bibinfo  {journal} {Phys. Rev. B}\ }\textbf {\bibinfo {volume} {41}},\
  \bibinfo {pages} {9323} (\bibinfo {year} {1990})}\BibitemShut {NoStop}%
\bibitem [{\citenamefont {H\"ormann}\ \emph {et~al.}(2018)\citenamefont
  {H\"ormann}, \citenamefont {Wunderlich},\ and\ \citenamefont
  {Schmidt}}]{Hoermann2018}%
  \BibitemOpen
  \bibfield  {author} {\bibinfo {author} {\bibfnamefont {M.}~\bibnamefont
  {H\"ormann}}, \bibinfo {author} {\bibfnamefont {P.}~\bibnamefont
  {Wunderlich}}, \ and\ \bibinfo {author} {\bibfnamefont {K.~P.}\ \bibnamefont
  {Schmidt}},\ }\href {\doibase 10.1103/PhysRevLett.121.167201} {\bibfield
  {journal} {\bibinfo  {journal} {Phys. Rev. Lett.}\ }\textbf {\bibinfo
  {volume} {121}},\ \bibinfo {pages} {167201} (\bibinfo {year}
  {2018})}\BibitemShut {NoStop}%
\bibitem [{\citenamefont {Koziol}\ \emph {et~al.}(2019)\citenamefont {Koziol},
  \citenamefont {Fey}, \citenamefont {Kapfer},\ and\ \citenamefont
  {Schmidt}}]{Koziol2019}%
  \BibitemOpen
  \bibfield  {author} {\bibinfo {author} {\bibfnamefont {J.}~\bibnamefont
  {Koziol}}, \bibinfo {author} {\bibfnamefont {S.}~\bibnamefont {Fey}},
  \bibinfo {author} {\bibfnamefont {S.~C.}\ \bibnamefont {Kapfer}}, \ and\
  \bibinfo {author} {\bibfnamefont {K.~P.}\ \bibnamefont {Schmidt}},\ }\href
  {\doibase 10.1103/PhysRevB.100.144411} {\bibfield  {journal} {\bibinfo
  {journal} {Phys. Rev. B}\ }\textbf {\bibinfo {volume} {100}},\ \bibinfo
  {pages} {144411} (\bibinfo {year} {2019})}\BibitemShut {NoStop}%
\bibitem [{\citenamefont {Knetter}\ and\ \citenamefont
  {Uhrig}(2000)}]{Knetter2000}%
  \BibitemOpen
  \bibfield  {author} {\bibinfo {author} {\bibfnamefont {C.}~\bibnamefont
  {Knetter}}\ and\ \bibinfo {author} {\bibfnamefont {G.~S.}\ \bibnamefont
  {Uhrig}},\ }\href {\doibase 10.1007/s100510050026} {\bibfield  {journal}
  {\bibinfo  {journal} {Eur. Phys. J. B}\ }\textbf {\bibinfo {volume} {13}},\
  \bibinfo {pages} {209} (\bibinfo {year} {2000})}\BibitemShut {NoStop}%
\bibitem [{\citenamefont {Knetter}\ \emph {et~al.}(2003)\citenamefont
  {Knetter}, \citenamefont {Schmidt},\ and\ \citenamefont
  {Uhrig}}]{Knetter2003}%
  \BibitemOpen
  \bibfield  {author} {\bibinfo {author} {\bibfnamefont {C.}~\bibnamefont
  {Knetter}}, \bibinfo {author} {\bibfnamefont {K.~P.}\ \bibnamefont
  {Schmidt}}, \ and\ \bibinfo {author} {\bibfnamefont {G.~S.}\ \bibnamefont
  {Uhrig}},\ }\href {\doibase 10.1088/0305-4470/36/29/302} {\bibfield
  {journal} {\bibinfo  {journal} {J. Phys. A}\ }\textbf {\bibinfo {volume}
  {36}},\ \bibinfo {pages} {7889} (\bibinfo {year} {2003})}\BibitemShut
  {NoStop}%
\bibitem [{\citenamefont {Coester}\ and\ \citenamefont
  {Schmidt}(2015)}]{Coester2015}%
  \BibitemOpen
  \bibfield  {author} {\bibinfo {author} {\bibfnamefont {K.}~\bibnamefont
  {Coester}}\ and\ \bibinfo {author} {\bibfnamefont {K.~P.}\ \bibnamefont
  {Schmidt}},\ }\href {\doibase 10.1103/PhysRevE.92.022118} {\bibfield
  {journal} {\bibinfo  {journal} {Phys. Rev. E}\ }\textbf {\bibinfo {volume}
  {92}},\ \bibinfo {pages} {022118} (\bibinfo {year} {2015})}\BibitemShut
  {NoStop}%
\bibitem [{Sup()}]{Suppl}%
  \BibitemOpen
  \href@noop {} {\bibinfo  {journal} {Details on the pCUT method, DlogPadé
  extrapolation techniques, linear spin-wave calculations and scaling
  relations, as well as multiplicative logarithmic exponents can be found in
  the Supplementary Material}\ }\BibitemShut {NoStop}%
\bibitem [{\citenamefont {Sachdev}(2011)}]{Sachdev2011}%
  \BibitemOpen
\bibfield  {journal} {  }\bibfield  {author} {\bibinfo {author} {\bibfnamefont
  {S.}~\bibnamefont {Sachdev}},\ }\href
  {https://books.google.de/books?id=F3IkpxwpqSgC} {\emph {\bibinfo {title}
  {Quantum Phase Transitions}}}\ (\bibinfo  {publisher} {Cambridge University
  Press},\ \bibinfo {year} {2011})\BibitemShut {NoStop}%
\end{thebibliography}%




\begin{thebibliography}
	\makeatletter
	\providecommand \@ifxundefined [1]{%
		\@ifx{#1\undefined}
	}%
	\providecommand \@ifnum [1]{%
		\ifnum #1\expandafter \@firstoftwo
		\else \expandafter \@secondoftwo
		\fi
	}%
	\providecommand \@ifx [1]{%
		\ifx #1\expandafter \@firstoftwo
		\else \expandafter \@secondoftwo
		\fi
	}%
	\providecommand \natexlab [1]{#1}%
	\providecommand \enquote  [1]{``#1''}%
	\providecommand \bibnamefont  [1]{#1}%
	\providecommand \bibfnamefont [1]{#1}%
	\providecommand \citenamefont [1]{#1}%
	\providecommand \href@noop [0]{\@secondoftwo}%
	\providecommand \href [0]{\begingroup \@sanitize@url \@href}%
	\providecommand \@href[1]{\@@startlink{#1}\@@href}%
	\providecommand \@@href[1]{\endgroup#1\@@endlink}%
	\providecommand \@sanitize@url [0]{\catcode `\\12\catcode `\$12\catcode
		`\&12\catcode `\#12\catcode `\^12\catcode `\_12\catcode `\%12\relax}%
	\providecommand \@@startlink[1]{}%
	\providecommand \@@endlink[0]{}%
	\providecommand \url  [0]{\begingroup\@sanitize@url \@url }%
	\providecommand \@url [1]{\endgroup\@href {#1}{\urlprefix }}%
	\providecommand \urlprefix  [0]{URL }%
	\providecommand \Eprint [0]{\href }%
	\providecommand \doibase [0]{http://dx.doi.org/}%
	\providecommand \selectlanguage [0]{\@gobble}%
	\providecommand \bibinfo  [0]{\@secondoftwo}%
	\providecommand \bibfield  [0]{\@secondoftwo}%
	\providecommand \translation [1]{[#1]}%
	\providecommand \BibitemOpen [0]{}%
	\providecommand \bibitemStop [0]{}%
	\providecommand \bibitemNoStop [0]{.\EOS\space}%
	\providecommand \EOS [0]{\spacefactor3000\relax}%
	\providecommand \BibitemShut  [1]{\csname bibitem#1\endcsname}%
	\let\auto@bib@innerbib\@empty
	
	\bibitem [{\citenamefont {Langheld}\ \emph {et~al.}(2022)\citenamefont
		{Langheld}, \citenamefont {Koziol}, \citenamefont {Adelhardt}, \citenamefont
		{Kapfer},\ and\ \citenamefont {Schmidt}}]{SLangheld2022}%
	\BibitemOpen
	\bibfield  {author} {\bibinfo {author} {\bibfnamefont {A.}~\bibnamefont
			{Langheld}}, \bibinfo {author} {\bibfnamefont {J.~A.}\ \bibnamefont
			{Koziol}}, \bibinfo {author} {\bibfnamefont {P.}~\bibnamefont {Adelhardt}},
		\bibinfo {author} {\bibfnamefont {S.~C.}\ \bibnamefont {Kapfer}}, \ and\
		\bibinfo {author} {\bibfnamefont {K.~P.}\ \bibnamefont {Schmidt}},\ }\href
	{\doibase 10.48550/ARXIV.2203.08081} {\enquote {\bibinfo {title} {Scaling at
				quantum phase transitions above the upper critical dimension},}\ } (\bibinfo
	{year} {2022})\BibitemShut {NoStop}%
	\bibitem [{\citenamefont {Knetter}\ and\ \citenamefont
		{Uhrig}(2000)}]{SKnetter2000}%
	\BibitemOpen
	\bibfield  {author} {\bibinfo {author} {\bibfnamefont {C.}~\bibnamefont
			{Knetter}}\ and\ \bibinfo {author} {\bibfnamefont {G.~S.}\ \bibnamefont
			{Uhrig}},\ }\href {\doibase 10.1007/s100510050026} {\bibfield  {journal}
		{\bibinfo  {journal} {Eur. Phys. J. B}\ }\textbf {\bibinfo {volume} {13}},\
		\bibinfo {pages} {209} (\bibinfo {year} {2000})}\BibitemShut {NoStop}%
	\bibitem [{\citenamefont {Knetter}\ \emph {et~al.}(2003)\citenamefont
		{Knetter}, \citenamefont {Schmidt},\ and\ \citenamefont
		{Uhrig}}]{SKnetter2003}%
	\BibitemOpen
	\bibfield  {author} {\bibinfo {author} {\bibfnamefont {C.}~\bibnamefont
			{Knetter}}, \bibinfo {author} {\bibfnamefont {K.~P.}\ \bibnamefont
			{Schmidt}}, \ and\ \bibinfo {author} {\bibfnamefont {G.~S.}\ \bibnamefont
			{Uhrig}},\ }\href {\doibase 10.1088/0305-4470/36/29/302} {\bibfield
		{journal} {\bibinfo  {journal} {J. Phys. A}\ }\textbf {\bibinfo {volume}
			{36}},\ \bibinfo {pages} {7889} (\bibinfo {year} {2003})}\BibitemShut 
	{NoStop}%
	\bibitem [{\citenamefont {Coester}\ and\ \citenamefont
		{Schmidt}(2015)}]{SCoester2015}%
	\BibitemOpen
	\bibfield  {author} {\bibinfo {author} {\bibfnamefont {K.}~\bibnamefont
			{Coester}}\ and\ \bibinfo {author} {\bibfnamefont {K.~P.}\ \bibnamefont
			{Schmidt}},\ }\href {\doibase 10.1103/PhysRevE.92.022118} {\bibfield
		{journal} {\bibinfo  {journal} {Phys. Rev. E}\ }\textbf {\bibinfo {volume}
			{92}},\ \bibinfo {pages} {022118} (\bibinfo {year} {2015})}\BibitemShut
	{NoStop}%
	\bibitem [{\citenamefont {Fey}\ \emph {et~al.}(2019)\citenamefont {Fey},
		\citenamefont {Kapfer},\ and\ \citenamefont {Schmidt}}]{SFey2019}%
	\BibitemOpen
	\bibfield  {author} {\bibinfo {author} {\bibfnamefont {S.}~\bibnamefont
			{Fey}}, \bibinfo {author} {\bibfnamefont {S.~C.}\ \bibnamefont {Kapfer}}, \
		and\ \bibinfo {author} {\bibfnamefont {K.~P.}\ \bibnamefont {Schmidt}},\
	}\href {\doibase 10.1103/PhysRevLett.122.017203} {\bibfield  {journal}
		{\bibinfo  {journal} {Phys. Rev. Lett.}\ }\textbf {\bibinfo {volume} {122}},\
		\bibinfo {pages} {017203} (\bibinfo {year} {2019})}\BibitemShut {NoStop}%
	\bibitem [{\citenamefont {Baker}(1975)}]{SBaker1975}%
	\BibitemOpen
	\bibfield  {author} {\bibinfo {author} {\bibfnamefont {G.}~\bibnamefont
			{Baker}},\ }\href@noop {} {\emph {\bibinfo {title} {Essentials of Pad{\'e}
				Approximants}}}\ (\bibinfo  {publisher} {Elsevier Science},\ \bibinfo {year}
	{1975})\BibitemShut {NoStop}%
	\bibitem [{\citenamefont {Guttmann}(1989)}]{SGuttmann1989}%
	\BibitemOpen
	\bibfield  {author} {\bibinfo {author} {\bibfnamefont {A.~J.}\ \bibnamefont
			{Guttmann}},\ }in\ \href@noop {} {\emph {\bibinfo {booktitle} {Phase
				Transitions and Critical Phenomena}}},\ Vol.~\bibinfo {volume} {13},\
	\bibinfo {editor} {edited by\ \bibinfo {editor} {\bibfnamefont
			{C.}~\bibnamefont {Domb}}, \bibinfo {editor} {\bibfnamefont {M.~S.}\
			\bibnamefont {Green}}, \ and\ \bibinfo {editor} {\bibfnamefont {J.~L.}\
			\bibnamefont {Lebowitz}}}\ (\bibinfo  {publisher} {Academic Press},\ \bibinfo
	{year} {1989})\BibitemShut {NoStop}%
	\bibitem [{\citenamefont {Xiao}(2009)}]{SXiao2009}%
	\BibitemOpen
	\bibfield  {author} {\bibinfo {author} {\bibfnamefont {M.-w.}\ \bibnamefont
			{Xiao}},\ }\href {\doibase 10.48550/ARXIV.0908.0787} {\enquote {\bibinfo
			{title} {Theory of transformation for the diagonalization of quadratic
				hamiltonians},}\ } (\bibinfo {year} {2009})\BibitemShut {NoStop}%
	\bibitem [{\citenamefont {Yusuf}\ \emph {et~al.}(2004)\citenamefont {Yusuf},
				\citenamefont {Joshi},\ and\ \citenamefont {Yang}}]{SYusuf2004}%
			\BibitemOpen
			\bibfield  {author} {\bibinfo {author} {\bibfnamefont {E.}~\bibnamefont
					{Yusuf}}, \bibinfo {author} {\bibfnamefont {A.}~\bibnamefont {Joshi}}, \ and\
				\bibinfo {author} {\bibfnamefont {K.}~\bibnamefont {Yang}},\ }\href {\doibase
				10.1103/PhysRevB.69.144412} {\bibfield  {journal} {\bibinfo  {journal} {Phys.
						Rev. B}\ }\textbf {\bibinfo {volume} {69}},\ \bibinfo {pages} {144412}
				(\bibinfo {year} {2004})}\BibitemShut {NoStop}%
	\bibitem [{\citenamefont {Fisher}(1974)}]{SFisher1974}%
	\BibitemOpen
	\bibfield  {author} {\bibinfo {author} {\bibfnamefont {M.~E.}\ \bibnamefont
			{Fisher}},\ }\href {\doibase 10.1103/RevModPhys.46.597} {\bibfield  {journal}
		{\bibinfo  {journal} {Rev. Mod. Phys.}\ }\textbf {\bibinfo {volume} {46}},\
		\bibinfo {pages} {597} (\bibinfo {year} {1974})}\BibitemShut {NoStop}%
	\bibitem [{\citenamefont {Fisher}(1983)}]{SFisher1983}%
	\BibitemOpen
	\bibfield  {author} {\bibinfo {author} {\bibfnamefont {M.~E.}\ \bibnamefont
			{Fisher}},\ }in\ \href@noop {} {\emph {\bibinfo {booktitle} {Critical
				Phenomena}}},\ \bibinfo {editor} {edited by\ \bibinfo {editor} {\bibfnamefont
			{F.~J.~W.}\ \bibnamefont {Hahne}}}\ (\bibinfo  {publisher} {Springer Berlin
		Heidelberg},\ \bibinfo {address} {Berlin, Heidelberg},\ \bibinfo {year}
	{1983})\ pp.\ \bibinfo {pages} {1--139}\BibitemShut {NoStop}%
	\bibitem [{\citenamefont {Binder}(1987)}]{SBinder1987}%
	\BibitemOpen
	\bibfield  {author} {\bibinfo {author} {\bibfnamefont {K.}~\bibnamefont
			{Binder}},\ }\href {\doibase 10.1080/00150198708227908} {\bibfield  {journal}
		{\bibinfo  {journal} {Ferroelectrics}\ }\textbf {\bibinfo {volume} {73}},\
		\bibinfo {pages} {43} (\bibinfo {year} {1987})}\BibitemShut {NoStop}%
	\bibitem [{\citenamefont {Berche}\ \emph {et~al.}(2012)\citenamefont {Berche},
		\citenamefont {Kenna},\ and\ \citenamefont {Walter}}]{SBerche2012}%
	\BibitemOpen
	\bibfield  {author} {\bibinfo {author} {\bibfnamefont {B.}~\bibnamefont
			{Berche}}, \bibinfo {author} {\bibfnamefont {R.}~\bibnamefont {Kenna}}, \
		and\ \bibinfo {author} {\bibfnamefont {J.-C.}\ \bibnamefont {Walter}},\
	}\href {\doibase https://doi.org/10.1016/j.nuclphysb.2012.07.021} {\bibfield
		{journal} {\bibinfo  {journal} {Nuclear Physics B}\ }\textbf {\bibinfo
			{volume} {865}},\ \bibinfo {pages} {115} (\bibinfo {year}
		{2012})}\BibitemShut {NoStop}%
	\bibitem [{\citenamefont {Kenna}\ and\ \citenamefont
		{Berche}(2013)}]{SKenna2013}%
	\BibitemOpen
	\bibfield  {author} {\bibinfo {author} {\bibnamefont {Kenna}}\ and\ \bibinfo
		{author} {\bibnamefont {Berche}},\ }\href {\doibase 10.5488/cmp.16.23601}
	{\bibfield  {journal} {\bibinfo  {journal} {Condensed Matter Physics}\
		}\textbf {\bibinfo {volume} {16}},\ \bibinfo {pages} {23601} (\bibinfo {year}
		{2013})}\BibitemShut {NoStop}%
	\bibitem [{\citenamefont {Sachdev}(2011)}]{SSachdev2011}%
	\BibitemOpen
	\bibfield  {journal} {  }\bibfield  {author} {\bibinfo {author} {\bibfnamefont
			{S.}~\bibnamefont {Sachdev}},\ }\href
	{https://books.google.de/books?id=F3IkpxwpqSgC} {\emph {\bibinfo {title}
			{Quantum Phase Transitions}}}\ (\bibinfo  {publisher} {Cambridge University
		Press},\ \bibinfo {year} {2011})\BibitemShut {NoStop}%
	\bibitem [{\citenamefont {SDutta}\ and\ \citenamefont
		{Bhattacharjee}(2001)}]{SDutta2001}%
	\BibitemOpen
	\bibfield  {author} {\bibinfo {author} {\bibfnamefont {A.}~\bibnamefont
			{Dutta}}\ and\ \bibinfo {author} {\bibfnamefont {J.~K.}\ \bibnamefont
			{Bhattacharjee}},\ }\href {\doibase 10.1103/PhysRevB.64.184106} {\bibfield
		{journal} {\bibinfo  {journal} {Phys. Rev. B}\ }\textbf {\bibinfo {volume}
			{64}},\ \bibinfo {pages} {184106} (\bibinfo {year} {2001})}\BibitemShut
	{NoStop}%
\end{thebibliography}


\bibliographystyle{apsrev4-1}

%%%%%%%%%%%%%%%%%%%%%%%%%%%%%%%%%%%%%%%%%%%%%%%%%%%%%%%%%%%%%%%%%%%%%%%%%%%%%%%%%%%%%%%%%%%% 
%%%%%%%%%% Merge with supplemental materials %%%%%%%%%%
%%%%%%%%%%%%%%%%%%%%%%%%%%%%%%%%%%%%%%%%%%%%%%%%%%%%%%%%%%%%%%%%%%%%%%%%%%%%%%%%%%%%%%%%%%%% 

\widetext
\clearpage
\begin{center}
	\textbf{\large Continuously varying critical exponents in long-range quantum spin ladders \\ \textit{Supplementary Material}} \\\vspace{0.5em}
	Patrick Adelhardt$^{\rm 1}$ and Kai Phillip Schmidt$^{\rm 1}$ \\
	\textit{$^{\text 1}$Friedrich-Alexander-Universit\"at Erlangen-N\"urnberg (FAU), \\Lehrstuhl f\"ur Theoretische Physik I, Staudtstra{\ss}e 7, D-91058 Erlangen, Germany}
\end{center}

%%%%%%%%%% Merge with supplemental materials %%%%%%%%%%
%%%%%%%%%% Prefix a "S" to all equations, figures, tables and reset the counter %%%%%%%%%%
\setcounter{equation}{0}
\setcounter{figure}{0}
\setcounter{table}{0}
\setcounter{page}{1}
\makeatletter
\renewcommand{\theequation}{S\arabic{equation}}
\renewcommand{\thefigure}{S\arabic{figure}}
\renewcommand{\bibnumfmt}[1]{[S#1]}
\renewcommand{\citenumfont}[1]{S#1}
%%%%%%%%%% Prefix a "S" to all equations, figures, tables and reset the counter %%%%%%%%%%

\thispagestyle{empty}

In the supplementary materials we first discuss the high-order series-expansion approach for long-range systems and extend the established approach to ground-state energies and observables. Second, we briefly introduce DlogPadé extrapolants and discuss the extrapolation scheme employed. Third, we provide the major steps of the linear spin-wave calculations used to determine the critical points within this approximations as a supplement to the pCUT results. Last, we provide an overview of scaling relations and hyperscaling which was only recently consistently generalized to the quantum long-range case \cite{SLangheld2022}. 

\section{High-order series expansion}

\subsection{The pCUT method}
In order to apply the pCUT method \cite{SKnetter2000, SKnetter2003} it must be possible to describe the problem under consideration with a Hamiltonian of the form
\begin{equation}
\mathcal{H} = \mathcal{H}_0 + \mathcal{V} = E_0 + \mathcal{Q} + \sum_{\delta > 0}^{\infty}\lambda(\delta) \mathcal{V}(\delta)\, 
\end{equation}
with an unperturbed Hamiltonian $\mathcal{H}_0$ with equidistant spectrum that is bounded from below and a perturbation $\mathcal{V}$. The unperturbed part becomes $\mathcal{H}_0 = E_0 + \mathcal{Q}$ with  $\mathcal{Q} = \sum_{i, \rho} t_{i, \rho}^{\dagger}t_{i, \rho}^{\phantom\dagger}$ counting the number of triplet quasiparticles (QPs). For long-range systems the perturbation $\mathcal{V}$ can be written as a sum between interacting processes of distance $\delta$ with a distance-dependent expansion parameter $\lambda(\delta)$. Also, the perturbation must decompose into
\begin{equation}
\mathcal{V} = \sum_{m=-N}^{N}\hat{T}_m =\sum_{m=-N}^{N}\sum_{l}\hat{\tau}_{m,l},
\end{equation}
where the operators $\hat{T}_m$ change the system's energy by $m$ energy quanta such that $[\mathcal{Q}, \hat{T}_m] = m \hat{T}_m$. The operator $\hat{T}_m$ decomposes into a sum of local operators $\hat{\tau}_{m,l}$ on a link $l$ connecting different sites of the underlying lattice.
When the above prerequisites are fulfilled the pCUT method unitarily transforms the original Hamiltonian, order by order in perturbation, to an effective, quasiparticle-conserving Hamiltonian $\mathcal{H}_{\text{eff}}$ reducing the complicated many-body problem to an easier effective few-body problem. The effective Hamiltonian in a generic form for an arbitrary number of expansion parameters $\lambda_i$ is then given by
\begin{equation}
\mathcal{H}_{\text{eff}} = \mathcal{H}_0 + \sum_{\sum_j n_j = k}^{\infty} \lambda_{1}^{n_1} \dots \lambda_{k}^{n_k} \sum_{\substack{\dim(\bm{m})=k, \\ \sum_i m_i=0}} C(\bm{m})\;\hat{T}_{m_1}\dots \hat{T}_{m_{k}}\, 
\label{Eq::h_eff}
\end{equation}
where the coefficients $C(\bm{m})$ are exactly given by rational numbers and the condition $\sum_i m_i=0$ enforces the quasiparticle conservation $[\mathcal{Q}, \mathcal{H}_{\text{eff}}]= 0$. Analogously, an effective observable is given by
\begin{equation}
\mathcal{O}_{\text{eff}} = \sum_{\sum_j n_j = k}^{\infty} \lambda_{1}^{n_1} \dots \lambda_{k}^{n_k} \sum_{i=1}^{k+1}\sum_{\dim(\bm{m})=k} \tilde{C}(\bm{m}; i)\;\hat{T}_{m_1}\dots\hat{T}_{m_{i-1}}\mathcal{O}\hat{T}_{m_i} \dots \hat{T}_{m_{k}}
\label{Eq::o_eff}
\end{equation}
with the rational coefficient $\tilde{C}(\bm{m}; i)$. In contrast to the effective Hamiltonian the effective observable is not quasiparticle conserving. The effective Hamiltonian and observables are generally independent of the exact form of the original Hamiltonian as long as the pCUT prerequisites are satisfied. To bring $\mathcal{H}_{\text{eff}}$ and $\mathcal{O}_{\text{eff}}$ into normal-ordered form, a model-dependent extraction process must be applied. For long-range interactions this is done most efficiently by a full-graph decomposition.

\subsection{Graph decomposition}

We apply the effective quantities to finite, topologically distinct graphs to bring them into normal-ordered structure. We refer to this approach as a linked-cluster expansion implemented as a full-graph decomposition. The underlying principle is the linked-cluster theorem which states that only linked processes have an overall contributions to cluster-additive quantities \cite{SCoester2015}. Since the effective pCUT Hamiltonian and observables are cluster-additive quantities we can reformulate Eqs.~\eqref{Eq::h_eff} and \eqref{Eq::o_eff} as
\begin{align}
\mathcal{H}_{\text{eff}} &= \mathcal{H}_0+ \sum_{\sum_j n_j = k}^{\infty} \lambda_{1}^{n_1} \dots \lambda_{k}^{n_k} \sum_{\substack{\dim(\bm{m})=k, \\ \sum_i m_i=0}}\sum_{\substack{\mathcal{G}, \\ |\mathcal{E}_{\mathcal{G}}|\le k}} C(\bm{m}) \sum_{\substack{l_1, \dots,l_k, \\ \bigcup_{i=1}^{k} l_i=\mathcal{G}}} \hat{\tau}_{m_1,l_1}\dots\hat{\tau}_{m_k, l_k}, 
\label{Eq::h_eff_linked} \\
\mathcal{O}_{\text{eff}} &= \sum_{\sum_j n_j = k}^{\infty} \lambda_{1}^{n_1} \dots \lambda_{k}^{n_k} \sum_{i=1}^{k+1} \sum_{\dim(\bm{m})=k}\sum_{\substack{\mathcal{G}, \\ |\mathcal{E}_{\mathcal{G}}|\le k}} \tilde{C}(\bm{m}; i) \sum_{\substack{l_1, \dots,l_k, \\ \bigcup_{i=1}^{k} l_i \cup x=\mathcal{G}}} \hat{\tau}_{m_1,l_1}\dots\hat{\tau}_{m_{i-1},l_{i-1}}\mathcal{O}_x\hat{\tau}_{m_i, l_i}\dots\hat{\tau}_{m_k, l_k},
\label{Eq::o_eff_linked}
\end{align}
where the sum over $\mathcal{G}$ runs over all possible simple connected graphs of perturbative order $k\ge |\mathcal{E}_{\mathcal{G}}|$. A graph $\mathcal{G}$ is a tuple $(\mathcal{E}_{\mathcal{G}}, \mathcal{V}_{\mathcal{G}})$ consisting of an edge or link set $\mathcal{E}_{\mathcal{G}}$ with $|\mathcal{E}_{\mathcal{G}}|$ edges and a set of vertices or sites $\mathcal{V}_{\mathcal{G}}$ with $|\mathcal{V}_{\mathcal{G}}|$ vertices. The conditions $\bigcup_{i=1}^{k} l_i=\mathcal{G}$ and $\bigcup_{i=1}^{k} l_i \cup x=\mathcal{G}$ arising from the linked-cluster theorem ensure that the cluster made up of active links and sites during a process must match with the edge and vertex set of a simple connected graph $\mathcal{G}$. Note, we generalized the notation for observables $\mathcal{O}_x$ where the index $x$ can either refer to a site (local observable) or a link (non-local observable). Thus, we can set up a full-graph decomposition applying the effective quantities to a set of finite, topologically distinct, simple connected graphs. \\
In the standard approach one would identify different expansion parameters with link colors which serve as another topological attribute in the classification of graphs. However, this approach fails for long-range interactions because every coupling parameter $\lambda(\delta)$ between sites of distance $\delta$ would be associated to a distinct link color and the number of graphs would already be infinite in first order of perturbation. We can overcome this obstacle by introducing white graphs \cite{SCoester2015} where different link colors are ignored in the topological classification of graphs and instead additional information is tracked during the calculation on white graphs. Only after the calculation during the embedding on the lattice the proper link color is introduced by reinserting the proper interaction strength. In particular, every link on a graph is associated with a distinct expansion parameter $\lambda_n^{\mathcal{G}}$ yielding a multivariable polynomial after applying the effective quantities to the graph. Only during the embedding on the lattice the expansion parameters of this polynomial is replaced by the actual coupling strength for each realization decaying algebraically with the distance between interacting sites. 

\subsection{Monte Carlo embedding}

Since we describe the ladder system in the language of rung dimers as super sites the graph contributions from the linked-cluster expansion must be embedded into a one-dimensional chain to determine the values of physical quantities $\kappa= \sum_m c_m^{(\kappa)}\lambda^m$ as a high-order series in the thermodynamic limit. Due to the infinite range of the algebraically decaying interactions every graph can be embedded infinitely many times at any order of perturbation. For each realization of a graph on the infinite chain the generic couplings $\lambda_n^{\mathcal{G}}$ in the multivariable polynomial corresponding to distinct edges is substituted by the true coupling strength $\lambda(-1)^{\delta}|\delta|^{-1-\sigma}$ or $\lambda(-1)^{1+\delta}|1+\delta|^{-1-\sigma}$ between graph vertices on sites $i$ and $i+\delta$ on the chain. For a prefactor $c_m$ in the high-order series only (reduced) contributions from graphs with up to $m$ links and $m+1$ sites can contribute. See Ref.~\cite{SCoester2015} for remarks about reduced quantities. We can write explicitly
\begin{equation}
c_m^{(\kappa)} = \sum_{N=2}^{m+1}  \sum_a f_N(a) = \sum_{N=2}^{m+1} S[f_N],
\end{equation}
where the first sum goes over the number of vertices and the second sum over all possible configurations excluding embeddings with overlapping vertices. The integrand $f_{N}$ combines all contributions from graphs with the same number of vertices $N$ since the $m-1$ sums contained in the sum $\sum_a$ are identical for graphs with the same number of vertices. The integration of these high-dimensional infinite nested sums $S[\cdot]$ quickly becomes very challenging when the perturbative order increases. It is essential to use Monte Carlo (MC) integration to evaluate these sums since MC techniques are known to be well suited for high-dimensional problems. We take a Markov-chain Monte Carlo approach to sample the configuration space \cite{SFey2019}. The fundamental moves consist of randomly selecting and moving graph vertices on the chain. For every embedding the integrands $f_N$ are evaluated with the correct couplings and added up to the overall contributions \cite{SFey2019}.

%\subsection{Application to physical quantities}

\section{DlogPadé extrapolations}

We extract the quantum-critical point including critical exponents from the pCUT method well beyond the radius of convergence of the pure high-order series using DlogPadé extrapolations. For a detailed description on DlogPadés and its application to critical phenomena we refer to Refs.~\cite{SBaker1975, SGuttmann1989}. The Padé extrapolant of a physical quantity $\kappa$ given as a perturbative series is defined as
\begin{equation}
P[L, M]_{\kappa} = \frac{P_L(\lambda)}{Q_{M}(\lambda)} = \frac{p_0 + p_1\lambda + \cdots + p_L\lambda^L }{1 + q_1\lambda + \cdots + q_M\lambda^M}
\label{Eq::Pade}
\end{equation}
with $p_i, q_i \in \mathbb{R}$ and the degrees $L$, $M$ of $P_{L}(x)$ and $Q_{M}(x)$ with $r\equiv L+M$, i.e., the Taylor expansion of Eq.~\eqref{Eq::Pade} about $\lambda=0$ up to order $r$ must recover the quantity $\kappa$ up to the same order. For DlogPadé extrapolants we introduce 
\begin{equation}
\mathcal{D}(\lambda) = \dv{\lambda}\ln(\kappa) \equiv P[L, M]_{\mathcal{D}}
\end{equation}
the Padé extrapolant of the logarithmic derivative $\mathcal{D}$ with $r-1=L+M$. Thus the DlogPadé extrapolant of $\kappa$ is given by
\begin{equation}
\mathrm{d}P[L,M]_{\kappa} = \exp\left(\int_0^{\lambda}P[L,M]_{\mathcal{D}} \,\mathrm{d}\lambda'\right).
\end{equation}
Given a dominant power-law behavior $\kappa \sim |\lambda - \lambda_c|^{-\theta}$, an estimate for the critical point $\lambda_c$ can be determined by excluding spurious extrapolants and analyzing the physical pole of $P[L,M]_{\mathcal{D}}$. If $\lambda_c$ is known, we can define biased DlogPadés by the Padé extrapolant 
\begin{equation}
\theta^{*} = (\lambda_c - \lambda)\dv{\lambda}\ln(\kappa) \equiv P[L,M]_{\theta^{*}}
\end{equation}
In the unbiased as well as the biased case we can extract estimates for the critical exponent $\theta$ by calculating the residua
\begin{equation}
\begin{split}
\theta_{\rm unbiased} =\Res P[L, M]_{\mathcal{D}}\vert_{\lambda=\lambda_c},\\
\theta_{\rm biased} = \Res P[L, M]_{\theta^{*}}\vert_{\lambda=\lambda_c}.
\end{split}
\end{equation}
At the upper critical dimension $\sigma=2/3$ multiplicative logarithmic corrections to the dominant power law behavior 
\begin{equation}
\kappa \sim \left|\lambda-\lambda_c \right|^{-\theta} \left(\ln\left(\lambda-\lambda_c\right)\right)^{p_{\theta}}
\end{equation}
in the vicinity of the quantum-critical point $\lambda_c$ are present. By biasing the critical point $\lambda_c$ and the exponent $\theta$ to its mean-field value, we define
\begin{equation}
p^{*}_{\theta} = -\ln(1-\lambda/\lambda_c)[(\lambda_c-\lambda)\mathcal{D}(\lambda)-\theta] \equiv P[L, M]_{p^*_{\theta}},
\end{equation}
such that we can determine an estimate for $p_{\theta}$ by again calculating the residuum of the Padé extrapolants $P[L,M]_{p^*_{\theta}}$. Note, for all quantities we calculate a large set of DlogPad\'e extrapolants with $L+M=r'\le r$, exclude defective extrapolants, and arrange the remaining DlogPad\'es in families with \mbox{$L-M=\text{const}$}. Although individual extrapolations deviate from each other, the quality of the extrapolations increases with the order of perturbation as members of different families but mutual order $r'$ converge. To systematically analyze the quantum-critical regime, we take the mean of the highest order extrapolants of different families with more than one member. %Multiplicative logarithmic exponents to the power law scaling for both ladder models $\mathcal{H}_{\shortparallel}$ and $\mathcal{H}_{\bowtie}$ can be found in Table~\ref{tab:MultiCorrections}.

%\begin{table}
%	\caption{\label{tab:MultiCorrections} Multiplicative logarithmic corrections $p_{\theta}$ at the upper critical dimension $\sigma=2/3$ associated to the ground-state energy $p_{\alpha}$, the elementary gap $p_{z\nu}$, and the 1QP spectral weight $p_{(2-z-\eta)\nu}$. Expected values from field-theoretical consideration are read of from Refs.~\cite{Wegner1973, Bauerschmidt2014}.}
%	\begin{ruledtabular}
%		\begin{tabular}{c|ccc}
%			\multirow{2}{*}{model} & \multicolumn{3}{c}{Multiplicative correction} \\  
%			& $p_{\alpha}$  & $p_{z\nu}$ & $p_{(2-z-\eta)\nu}$  \\
%			\hline
%			Field-theoretical predictions& $-\frac{5}{22}\approx-0.227$ & $\frac{1}{11}\approx0.091$ & $???$\\
%			$\mathcal{H}_{\shortparallel}$ & & & \\
%			$\mathcal{H}_{\bowtie}$ & & & \\		
%		\end{tabular}
%	\end{ruledtabular}
%\end{table}

\section{Linear Spin-wave calculations}
We supplement the critical behavior determined by the pCUT approach with critical points from linear spin-wave approximation. As spin-wave theory considers fluctuations about the classical ground state it is certainly valid in the Néel-ordered phase of the long-range Heisenberg ladders. We start by mapping the spin operators to boson creation and annihilation operators using the Holstein-Primakoff transformation up to linear order in the boson operators. For the antiferromagnetic Heisenberg spin ladder the system must be divided into two sublattices constituting the expected antiferromagnetic Néel order for strong long-range interactions. The transformation thus reads
\begin{equation}
\begin{gathered}
\begin{aligned}
S_{i,1}^{z}&= S - a_{i,1}^{\dagger}a_{i,1}^{\phantom\dagger} &\quad S_{i,1}^{-}&\approx \sqrt{2S}a_{i, 1}^{\dagger} & \quad S_{i,1}^{+}&\approx\sqrt{2S}a_{i,1}^{\phantom\dagger},
\end{aligned}\\
\begin{aligned}
S_{i,2}^{z}&= b_{i,2}^{\dagger}b_{i,2}^{\phantom\dagger} - S &\quad S_{i,2}^{-}&\approx \sqrt{2S}b_{i, 2}^{\phantom\dagger} & \quad S_{i,2}^{+}&\approx\sqrt{2S}b_{i,2}^{\dagger},
\end{aligned}\\
\begin{aligned}
S_{j,1}^{z}&= b_{j,1}^{\dagger}b_{j,1}^{\phantom\dagger} - S &\quad S_{j,1}^{-}&\approx \sqrt{2S}b_{j,1}^{\phantom\dagger} & \quad S_{j,1}^{+}&\approx\sqrt{2S}b_{j,1}^{\dagger},
\end{aligned}\\
\begin{aligned}
S_{j,2}^{z}&= S - a_{j,2}^{\dagger}a_{j,2}^{\phantom\dagger} &\quad S_{j,2}^{-}&\approx \sqrt{2S}a_{j,2}^{\dagger} & \quad S_{j,2}^{+}&\approx\sqrt{2S}a_{j,2}^{\phantom\dagger}
\end{aligned}
\end{gathered}
\end{equation}
with $i$ odd and $j$ even rungs. Inserting these identities into the Hamiltonian $\mathcal{H}_{\shortparallel}$, neglecting quartic terms and Fourier transforming the problem we arrive at
\begin{equation}
\begin{split}
\mathcal{H}_{\shortparallel}^{\text{SW}} \approx \text{const.} + S \sum_{k} \Big\{&\sum_{\nu}\left[\left(\gamma - f(k)\right)\left(a_{k, \nu}^{\dagger}a_{k, \nu}^{\phantom\dagger} + b_{-k, \nu}^{\dagger}b_{-k, \nu}^{\phantom\dagger} \right) + g(k)\left(a_{k, \nu}^{\phantom\dagger}b_{-k, \nu}^{\phantom\dagger} + a_{k, \nu}^{\dagger}b_{-k, \nu}^{\dagger} \right)\right] \\ &+ a_{k, 1}^{\phantom\dagger}b_{-k, 2}^{\phantom\dagger} +  a_{k, 2}^{\phantom\dagger}b_{-k, 1}^{\phantom\dagger} + a_{k, 1}^{\dagger}b_{-k, 2}^{\dagger} + a_{k, 2}^{\dagger}b_{-k, 1}^{\dagger} \Big\}.
\end{split}
\end{equation}
Incorporating the long-range couplings for an infinite chain into the prefactors we can define the quantities
\begin{equation}
\begin{gathered}
\begin{aligned}
&\gamma = 1 + 2\lambda \sum_{\delta = 1}^{\infty} \frac{1}{(2\delta - 1)^{1+\sigma}}, \qquad &f(k)= 2\lambda \sum_{\delta=1}^{\infty} \frac{\cos(2k\delta)-1}{(2\delta)^{1+\sigma}}, \qquad &g(k) = 2\lambda \sum_{\delta=1}^{\infty}\frac{\cos\left[(2\delta-1)k\right]}{(2\delta - 1)^{1+\sigma}}.
\end{aligned}
\end{gathered}
\end{equation}
This Hamiltonian is quadratic in creation and annihilation operators in quasimomenta and we intend to diagonalize the problem employing a Bogoliubov-Valatin transformation. Following Ref.~\cite{SXiao2009} we introduce the operator
\begin{equation}
\vec{\psi}_{k}^{\dagger} = \begin{pmatrix} \vec{c}_{k}^{\dagger} & \vec{c}_{k}^{T} \end{pmatrix} = \begin{pmatrix} a_{k,1}^{\dagger} & b_{-k,1}^{\dagger} & a_{k,2}^{\dagger} & b_{-k,2}^{\dagger} & a_{k,1}^{\phantom\dagger} & b_{-k,1}^{\phantom\dagger} & a_{k,2}^{\phantom\dagger} & b_{-k,2}^{\phantom\dagger}	\end{pmatrix}.
\end{equation}
We use this operator to bring the spin-wave Hamiltonian into canonical quadratic form
\begin{equation}
\mathcal{H}_{\shortparallel}^{\text{SW}} = \sum_{k} \Bigg[ \frac{1}{2} \vec{\psi}^{\dagger}\underbrace{\begin{pmatrix} A_k^{\phantom\dagger} & B_k^{\phantom\dagger} \\ B_k^{\dagger} & A_k^{T} \end{pmatrix}}_{\equiv M_k} \vec{\psi}^{\phantom\dagger} - \frac{1}{2}\tr A_k^{\phantom\dagger} \Bigg],
\end{equation}
where $A_k$ and $M_k$ are Hermitian matrices and $B_k$ is a symmetric matrix. To solve the diagonalization problem we must find a transformation $\vec{\psi}_k = T\vec{\varphi}_k$ that brings $M_k$ into diagonal form and preserves the bosonic anticommutation relations of $\vec{\psi}_k$. Xiao \cite{SXiao2009} proofs that the problem can be reformulated in terms of the eigenvalue problem of the dynamic matrix
\begin{equation}
D_k = \begin{pmatrix} A_k^{\phantom\dagger} & B_k^{\phantom\dagger} \\ -B_k^{\dagger} & -A_k^{T} \end{pmatrix}
\end{equation}
arising from the Heisenberg equation of motion and that the transformation matrix $T$ can be constructed using appropriately normalized eigenvectors. A physical solution to the problem exists if and only if the dynamical matrix is diagonalizable and the eigenvalues are real. Employing this scheme we find
\begin{equation}
\mathcal{H}_{\shortparallel}^{\text{SW}} = \text{const.} + S \sum_{k,\nu}\left(\omega_{+}(k) \alpha_{k,\nu}^{\dagger}\alpha_{k,\nu}^{\phantom\dagger} + \omega_{-}(k)\beta_{k,\nu}^{\dagger}\beta_{k,\nu}^{\phantom\dagger}\right)
\end{equation}
in terms of the new boson creation and annihilation operators $\alpha_{k,\nu}^{(\dagger)} $ and $\beta_{k,\nu}^{(\dagger)}$ and the spin-wave dispersion
\begin{equation}
\omega_{\pm}(k) = \sqrt{\left(\gamma - f(k)\right)^2 - \left(g(k) \pm 1\right)^2}.
\end{equation}
In the limit $\lambda\rightarrow\infty$ we recover the spin-wave dispersion in Ref.~\cite{SYusuf2004} for the long-range Heisenberg spin chain. The staggered magnetization deep in the antiferromagnetic regime can be expressed as $m =  S - \Delta m$ where $\Delta m$ is the correction induced by quantum fluctuations. We start with the expression
\begin{equation}
\Delta m = \sum_{\nu=1}^{2} \langle a_{j,\nu}^{\dagger}a_{j,\nu}\rangle \overset{N\rightarrow \infty}{=} \frac{1}{\pi} \sum_{\nu}^{2} \int_{-\pi/2}^{\pi/2}\mathrm{d}k  \langle  a_{k,\nu}^{\dagger}a_{k,\nu} \rangle
\end{equation}
and rewriting it in terms of the boson operators $\alpha_{k,\nu}^{(\dagger)} $ and $\beta_{k,\nu}^{(\dagger)}$ we find
\begin{equation}
\Delta m = \frac{1}{\pi} \int_{-\pi/2}^{\pi/2} \mathrm{d}k \left[\frac{1}{2}\left(\frac{\gamma - f(k)}{\omega_{+}(k)} + \frac{\gamma - f(k)}{\omega_{-}(k)}\right) - 1\right].
\end{equation}
Introducing the linear Holstein-Primakoff transformation for the Hamiltonian $\mathcal{H}_{\bowtie}$ including diagonal long-range interactions the linear spin-wave Hamiltonian reads
\begin{equation}
\begin{split}
\mathcal{H}_{\bowtie}^{\text{SW}} = \text{const.} + S \sum_{k} \Big\{&\sum_{\nu}\left[\left(\Gamma - f(k)\right)\left(a_{k, \nu}^{\dagger}a_{k, \nu}^{\phantom\dagger} + b_{-k, \nu}^{\dagger}b_{-k, \nu}^{\phantom\dagger} \right) + g(k)\left(a_{k, \nu}^{\phantom\dagger}b_{-k, \nu}^{\phantom\dagger} + a_{k, \nu}^{\dagger}b_{-k, \nu}^{\dagger} \right)\right] \\ &+ v(k)\left(a_{k, 1}^{\phantom\dagger}b_{-k, 2}^{\phantom\dagger} +  a_{k, 2}^{\phantom\dagger}b_{-k, 1}^{\phantom\dagger} + a_{k, 1}^{\dagger}b_{-k, 2}^{\dagger} + a_{k, 2}^{\dagger}b_{-k, 1}^{\dagger}\right) \\
&+ w(k)\left(a_{k,1}^{\dagger}a_{k,2}^{\phantom\dagger} + a_{k,2}^{\dagger}a_{k,1}^{\phantom\dagger} + b_{-k,1}^{\dagger}b_{-k,2}^{\phantom\dagger} + b_{-k,2}^{\dagger}b_{-k,1}^{\phantom\dagger} \right)\Big\},
\end{split}
\end{equation}
where we introduced the multiple prefactors defined as $\kappa = \kappa_1 + \kappa_2$, $\Gamma = \gamma + \kappa$ and as
\begin{equation}
\begin{gathered}
\begin{aligned}
%&\gamma = 1 + 2\lambda \sum_{\delta = 1}^{\infty} \frac{1}{(2\delta - 1)^{1+\sigma}}, \quad &f(k)= 2\lambda \sum_{\delta=1}^{\infty} \frac{\cos(2k\delta)-1}{(2\delta)^{1+\sigma}} \quad &g(k) = 2\lambda \sum_{\delta=1}^{\infty}\frac{\cos\left[(2\delta-1)k\right]}{(2\delta - 1)^{1+\sigma}} \\
&\kappa_1 =  2\lambda \sum_{\delta=1}^{\infty}\frac{1}{\left((2\delta)^2+1\right)^{\frac{1+\sigma}{2}}}, \quad &\kappa_2 = 2\lambda \sum_{\delta=1}^{\infty} \frac{1}{\left((2\delta-1)^2+1\right)^{\frac{1+\sigma}{2}}}, \quad &v(k) = 1 + 2\lambda \sum_{\delta=1}^{\infty} \frac{\cos(2\delta k)}{\left((2\delta)^2+1\right)^{\frac{1+\sigma}{2}}}, \\
&&w(k)= 2\lambda \sum_{\delta=1}^{\infty} \frac{\cos\left[(2\delta-1)k\right]}{\left((2\delta-1)^2+1\right)^{\frac{1+\sigma}{2}}}.
\end{aligned}
\end{gathered}
\end{equation}
Again employing the same Bogoliubov-Valatin transformation we can derive the spin-wave dispersion
\begin{equation}
\omega_{\pm}(k) = \sqrt{\left[\Gamma - (f(k)\pm w(k))\right]^2 - \left[g(k)\pm v(k)\right]^2}
\end{equation}
and the corrections to the staggered magnetization
\begin{equation}
\Delta m = \frac{1}{\pi} \int_{-\pi/2}^{\pi/2} \mathrm{d}k \left[\frac{1}{2}\left(\frac{\Gamma - f(k) - w(k)}{\omega_{+}(k)} + \frac{\Gamma - f(k) + w(k)}{\omega_{-}(k)}\right) - 1\right].
\end{equation}
For both Hamiltonians $\mathcal{H}_{\shortparallel}$ and $\mathcal{H}_{\bowtie}$ we evaluate the integrals $\Delta m$ numerically and use the consistency condition $\Delta m < S$ in the antiferromagnetic regime to approximate the phase transition point.

\section{(Hyper-)scaling relations}
In RG theory the generalized homogeneity of the free energy density is exploited \cite{SFisher1974}. Connecting the critical exponents of observables with the derivatives of the free energy density and exploiting the homogeneity properties the \mbox{(hyper-)} scaling relations
\begin{align}
\gamma &= (2-\eta)\nu \qquad~\text{(Fisher equality)}, \label{eq:fisher}\\
\gamma &= \beta(\delta - 1) \qquad~\text{(Widom equality)}, \label{eq:widom}\\
2 &= \alpha + 2\beta + \gamma \quad\,\text{(Essam-Fisher equality)}, \label{eq:essam}\\
2-\alpha &= \left(d + z\right)\nu \quad\quad\,\text{(Hyperscaling relation)}
\end{align}
can be derived. However, the hyperscaling relation breaks down above the upper critical dimension due to dangerous irrelevant variables in the free energy sector since these variables cannot be set to zero as the free energy density becomes singular in this limit \cite{SFisher1983, SBinder1987}. Allowing the correlation sector to be affected by dangerous irrelevant variables for quantum systems in analogy to previous works in classical systems \cite{SBerche2012, SKenna2013} the hyperscaling relation can be generalized to
\begin{equation}
2-\alpha = \left(\frac{d}{\koppa} + z\right)\nu \label{eq:gen_hyperscaling}
\end{equation} 
with the pseudocritical exponent $\koppa=\max\left(1, d/d_{\text{uc}}\right)$ \cite{SLangheld2022}. As the one-dimensional $O(3)$ quantum rotor model can be mapped to the low-energy properties of the dimerized antiferromagnetic Heisenberg ladder \cite{SSachdev2011} we can insert the long-range mean-field critical exponents 
\begin{equation}
\begin{gathered}
\begin{aligned}
&\gamma = 1, \qquad &\nu = \frac{1}{\sigma}, &\qquad z = \frac{\sigma}{2}, \qquad &\eta= 2 - \sigma
\end{aligned}
\end{gathered}
\end{equation}
derived from one-loop RG \cite{SDutta2001} for quantum rotor models at the upper critical dimension into Eq.~\eqref{eq:essam}. We find $d_{\text{uc}}(\sigma)=3\sigma/2$. It directly follows that $d>d_{\text{uc}}$ in the regime $\sigma < 2/3$. Thus, we can rewrite
\begin{equation}
\koppa = \max\left( 1, \frac{2}{3\sigma} \right) = 
\begin{cases}  
1 & \text{ for } \sigma \geq 2/3\\
\frac{2}{3\sigma} & \text{ for } \sigma < 2/3.
\end{cases}
\end{equation}
which together with equation Eq.~\eqref{eq:gen_hyperscaling} is the generalized hyperscaling relation as derived in Ref.~\cite{SLangheld2022}.

\bibliographystyle{apsrev4-1}

\end{document}